\documentclass[aps,pra,reprint,superscriptaddress]{revtex4-1}

\usepackage{fancyvrb}
\usepackage{amsmath}
\usepackage{amssymb}
\usepackage{hyperref}
\usepackage{graphicx}
\usepackage{mathtools}
\usepackage[T1]{fontenc}
\usepackage{braket}
\usepackage{mathrsfs}
\usepackage[caption=false]{subfig}
\usepackage{color}
\newcommand\dif{\mathop{}\!\mathrm{d}}

\begin{document}


\title{Efficient quantum transport in disordered Floquet networks}


\author{Hl\'er Kristj\'ansson}
\affiliation{Physikalisches Institut, Albert-Ludwigs-Universit\"at Freiburg, Hermann-Herder-Stra{\ss}e 3, D-79104 Freiburg, Germany}
\affiliation{Department of Physics, Blackett Laboratory, Imperial College London, SW7 2AZ, London, United Kingdom}
\affiliation{D\'epartement d'informatique et de recherche op\'erationnelle \& Institut Courtois, Universit{\' e} de Montr{\' e}al, H3T 1J4, Montr{\' e}al, Qu{\' e}bec,  Canada}

\author{Jonathan Brugger}
\affiliation{Physikalisches Institut, Albert-Ludwigs-Universit\"at Freiburg, Hermann-Herder-Stra{\ss}e 3, D-79104 Freiburg, Germany}

\author{Gabriel Dufour}
\affiliation{Physikalisches Institut, Albert-Ludwigs-Universit\"at Freiburg, Hermann-Herder-Stra{\ss}e 3, D-79104 Freiburg, Germany}
\affiliation{Freiburg Institute for Advanced Studies, Albert-Ludwigs-Universit{\"a}t-Freiburg, Albertstra{\ss}e 19, D-79104 Freiburg, Germany}

\author{Christian Scheppach}
\affiliation{Physikalisches Institut, Albert-Ludwigs-Universit\"at Freiburg, Hermann-Herder-Stra{\ss}e 3, D-79104 Freiburg, Germany}

\author{Andreas Buchleitner}
\affiliation{Physikalisches Institut, Albert-Ludwigs-Universit\"at Freiburg, Hermann-Herder-Stra{\ss}e 3, D-79104 Freiburg, Germany}
\email{Correspondence: abu@uni-freiburg.de} 
 


\begin{abstract}
We propose a mechanism for fast and efficient quantum transport through disordered networks with variable on-site energies, inspired by photosynthetic complexes. The mechanism relies on an interplay between inter-site couplings of the network and driving by external vibrations. Two design principles are shown to ensure close-to-perfect transport despite the disorder, namely a reflection symmetry in Floquet-Hilbert space and the existence of a dominant doublet or triplet of Floquet states. 

\end{abstract}


\maketitle

Fast and efficient transfer of quantum states across a network is a subject of great importance in fields ranging from quantum computing to biological light-harvesting. Fine-tuning quantum systems to allow for perfect state transfer from one location to another has been of considerable interest in the quantum information community \cite{Bose03,Christandl04, Keil13, Wojcik07}. On the other hand, disorder is present in any realistic system, and while it is generally thought to hinder quantum transport \cite{Anderson58}, it has been shown that many quantum information tasks can equally be performed in systems with a high degree of disorder \cite{Chakraborty16}.
Moreover, disorder can also be viewed as a statistical tool to generate networks which allow for efficient transfer, without control over individual parameters \cite{Scholak10, Scholak11, Walschaers13, Walschaers15, Mohseni13}. In this approach, minimal assumptions on the fluctuating parameters are shown to dramatically enhance the probability of obtaining realizations with the desired transport properties with respect to the fully random case \cite{Walschaers13,Walschaers15}. Such \emph{design principles} therefore provide a robust way of ensuring good transfer in otherwise poorly controlled systems.

In this work, we investigate excitation transport through a network of two-level systems (`sites')
whose excitation energies decrease from the input to the output of the network. We suppose that all sites are coupled, with coupling strengths consisting of a constant term plus a small harmonic modulation. Both the static couplings and the oscillation amplitudes are randomly drawn from a  Gaussian distribution.
Because of the large difference in excitation energies of the input and output sites, an excitation initially localized at the input site has negligible probability of being transported to the output site in a static network. However, the small periodic modulation of the inter-site couplings can enable perfect transport if the oscillation frequency bridges the energy gap between the input and output sites. This resonance is a necessary but not sufficient condition for near-perfect quantum transport.

  An inspiration for this setup is experimental data from biological light-harvesting systems \cite{SchmidtAmBusch11}, such as the FMO photosynthetic complex \cite{Blankenship02}, where potential quantum coherence effects have been reported \cite{Engel07,Christensson12,Ishizaki09,Scholes17,Dostal16} (the extent of which is the subject of current debate \cite{duan2017nature,cao2020quantum,zerah2021photosynthetic,higgins2021photosynthesis,mattioni2021design,lorenzoni2025full}). The FMO complex transports excitations captured by a light-harvesting antenna across a network of dipole-coupled chlorophyll molecules to a reaction center in green sulfur bacteria.
  However, evidence in \cite{SchmidtAmBusch11} suggests that the transition energies of the molecules decrease across the network, forbidding resonant transfer between the input and output sites. Efficient energy transfer therefore requires coupling to external degrees of freedom. In particular, recent experimental and theoretical work on the FMO photosynthetic complex underlines the importance of the interplay between vibrational and electronic degrees of freedom to allow for energy transport in such systems \cite{Christensson12, Aghtar14, Irish14, Nalbach15,gorman2018engineering,li2022interplay,policht2022hidden,higgins2021photosynthesis}.

We propose two design principles for the ensemble of Hamiltonians describing the disordered networks driven on resonance, which are sufficient for near-perfect quantum transport. 
The evolution of the periodically driven system is described using Floquet theory \cite{Floquet,Shirley65,Zeldovich67,oka2019floquet}, and the design principles are antisymmetry of the Hamiltonian in Floquet-Hilbert space and the existence of a dominant doublet (or triplet)  of Floquet states, which collectively induce behavior reminiscent of transport in a two- (or three-) site system. The presence of the intermediate sites is, however, shown to allow transport on timescales much faster than that which would occur through direct coupling between the input and output sites only. Such transport which is both fast and efficient, is essential for excitation or information transfer in any physical network subject to decoherence or loss.

\emph{The model.---} We consider a network of $N$ two-level systems and work in the single-excitation Hilbert space $\mathcal{H}$ with basis vectors $\ket{n}$, $n \in \{1,\dots ,N\}$, corresponding to the excitation localized on site $n$.
To begin with, we consider the Hamiltonian 
\begin{equation}
\label{eq:H_0}
H_0 =  \sum_{n,m=1}^{N} V_{nm}\ket{n}\bra{m},
\end{equation}
with $V_{nn}$ the excitation energy of site $n$ and $V_{nm}$ the (static) coupling between sites $n$ and $m$.
To account for the variability between realizations of the network, $H_0$ is sampled from the Gaussian orthogonal ensemble (GOE) \cite{Walschaers15,mehta2004random} with variance $\sigma_0^2$ away from the diagonal, and $2 \sigma_0^2$ on the diagonal.

\begin{figure}
	\centering
	\includegraphics[width=\linewidth]{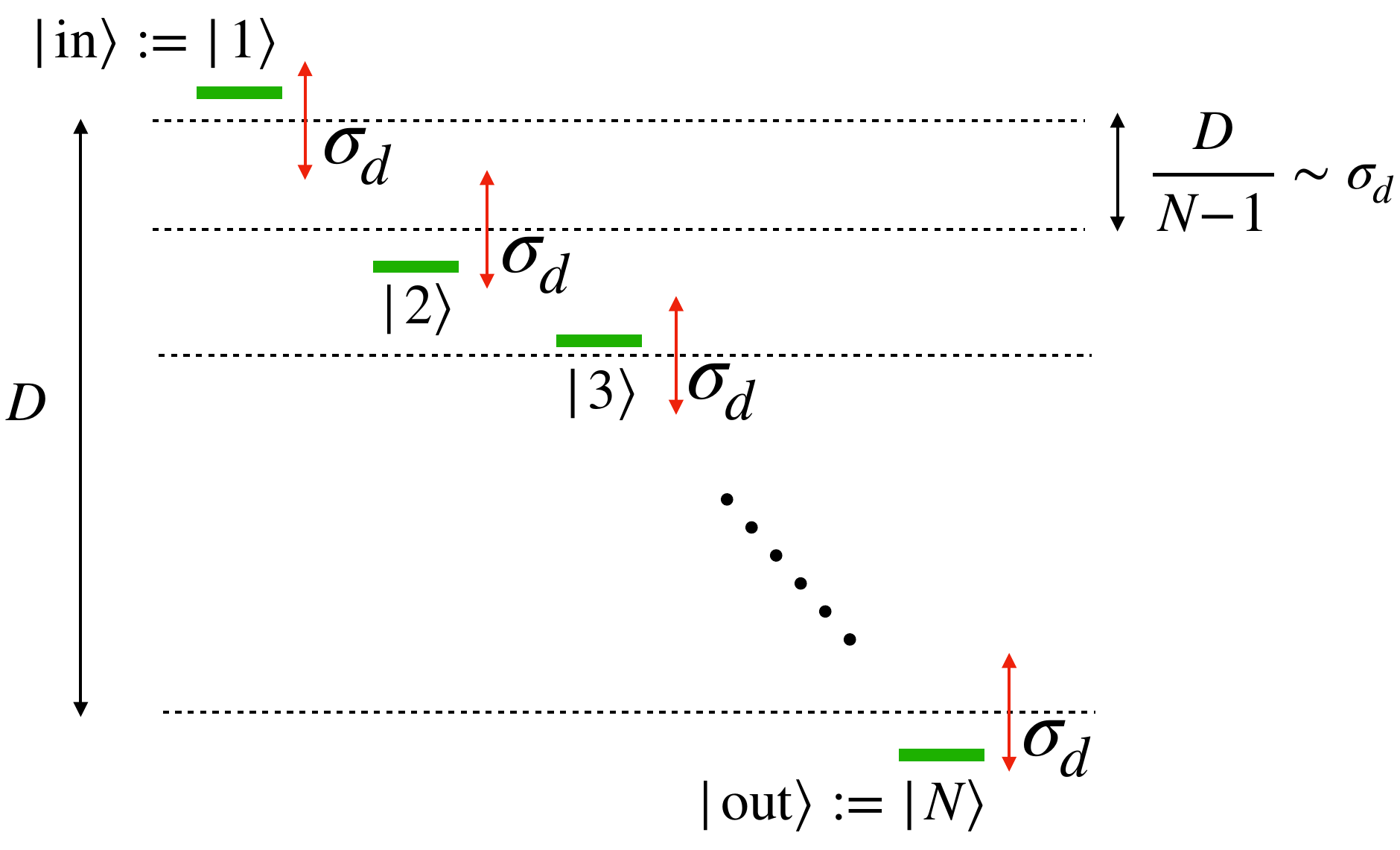}%
	\caption{\label{fig:diagram} 
Diagram of the energy levels of the diagonal $N$-site Hamiltonian $H_\mathrm{d}$. The energies corresponding to each site are normally distributed with mean $D/2 - D(n-1)/(N-1 )$ and variance $\sigma_d^2$, where $\sigma_d$ is taken to be of comparable size as the average energy difference $D/N$ between neighboring sites. Moreover, we take $\sigma_d=\sigma_0$, the standard deviation of the couplings of the static Hamiltonian $H_0$.  
}
\end{figure}

Note, that in the Hamiltonian $H_0$, there is no sense of direction across the network or of proximity between sites. This lack of directionality is broken by adding a diagonal term $H_\mathrm{d}=\sum_{n=1}^{N}d_n \ket{n}\bra{n}$ to the Hamiltonian, to account for the energy detuning between sites. Here each $d_{n}$ is normally distributed with mean $D/2 - D(n-1)/(N-1 )$, where $D$ is the mean energy difference between input and output sites, 
and variance $\sigma_\mathrm{d}^2 = \sigma_0^2$ (see Figure \ref{fig:diagram}). This energy difference  is taken to be much larger than the typical coupling strength, i.e.~$D \gg \sigma_0$, while the detuning between neighboring sites is taken to be comparable to their coupling, i.e.~$D/N \sim \sigma_0$.

This singles out sites $n=1$ and $n=N$, which have (on average) the highest and lowest excitation energies and which we identify as the input and  output sites of the network, respectively. The other sites form an irregular energy ladder between the two. 
With this choice, the spectral decomposition of the static Hamiltonian $H_0+H_\mathrm{d}=\sum_n e_n \ket{e_n}\bra{e_n}$ (with $e_1 \ge e_2 \ge \dots \ge e_N$ and $\braket{n|e_n}$ taken to be real and positive) yields eigenstates $\ket{e_n}$ which are predominantly located on site $n$ and its neighbors but no fully delocalized excitonic states that could mediate transport across the network.

We now assume that the network is coupled to a vibrational mode of frequency $\omega$ tuned to the transition frequency $\omega_0=e_1-e_N$ between the highest and lowest energy eigenstates of the static Hamiltonian.
 The transfer from input to output of the network can thus be realized by emitting a single phonon. In other words, vibrations dress the input state of the network, bringing it in resonance with the output state \cite{CT08}. Note that while many other vibrational modes will typically be present in realistic systems, we assume that this resonant process is dominant. Later, we also consider the case where resonance is achieved by emitting two phonons of frequency $\omega_0/2$.
 The vibrations result in small time-periodic modulations of the on-site energies and inter-site couplings, with the amplitudes of the oscillations gathered in a matrix $H_\mathrm{v} =  \sum_{n,m=1}^{N} W_{nm}\ket{n}\bra{m}$ which is  randomly sampled from the GOE with variance $\sigma_\mathrm{v}^2 \ll \sigma_0^2$. 
The full Hamiltonian is thus given by
\begin{equation}
\label{eq:full_ham}
H (  t  )   =      H_0+H_\mathrm{d} + \sin(  \omega t  ) H_\mathrm{v} \, .
\end{equation}

The system is initialized at time $t_0$ with the excitation localized on the input site,  i.e. $\ket{\psi(t = t_0)} = \ket{\textrm{in}} := \ket{\textrm{1}}$.
Given that we have fixed the phase of the oscillations in Eq.\ \eqref{eq:full_ham}, the initial time $t_0$ needs to be explicitly taken into consideration. However, with our choice of parameters, the choice of $t_0$ is found not to impact the overall transport behavior. In the numerical simulations, a random initial time $t_0, ~ \left(0 \leq  t_0 < T :=2\pi/\omega \right)$ is chosen for each realization. 
The \textit{transfer probability} at time $t$ is then given by the probability of finding the  excitation on the output site $\ket{\textrm{out}} := \ket{N}$:
	\begin{equation}
	\label{eq:P(t)}
	P(t) := |\braket{\textrm{out}|\psi(t)}|^2.
	\end{equation}
Transport is said to be efficient if $P(t)$ reaches a maximum $P_\mathrm{max}$ close to one. In real systems, this should occur within the timescale defined by the onset of decoherence and loss.
Note that we do not consider any in- or out-coupling mechanism for the excitation, therefore the dynamics is unitary and $P(t)$ will typically display an oscillatory behavior as the excitation travels back and forth across the network. In the following, we only consider the first such oscillation.

\emph{Design principles.---} 
Transport across a static network described by $H_0$ alone (see Eq.~\eqref{eq:H_0}) was extensively studied in Refs.\ \cite{Walschaers13,Walschaers15,Walschaers16}. There, it was shown that a transfer probability close to one is ensured if the dynamics is dominated by a pair of eigenstates resembling the tunneling doublet of a symmetric double-well. Moreover,  \textit{centrosymmetry} of the Hamiltonian was found to enhance the probability of obtaining a realization displaying such a \textit{dominant doublet} \cite{Zech14,Walschaers13}. A Hamiltonian is centrosymmetric if it commutes with the exchange operator $J := \sum_{n=1}^N \ket{N-n+1}\bra{n}$, i.e.~if the network exhibits a reflection symmetry around its center. This in particular ensures that the input and output sites have equal excitation energies, allowing for resonant transfer between the two.
In this model, the addition of the on-site energy gradient $H_\mathrm{d}$ is incompatible with  centrosymmetry and precludes 
the existence of a dominant doublet in the static Hamiltonian $H_0+H_\mathrm{d}$.
In the present work, we show how the introduction of periodic driving allows to recover conceptually similar design principles.

The time-periodic Hamiltonian in Eq.\ \eqref{eq:full_ham} is conveniently treated using Floquet theory \cite{Floquet,Shirley65,Zeldovich67,oka2019floquet}. For a system evolving under a time-dependent Hamiltonian $H(t)$ with period $T=2\pi/\omega$, the Floquet theorem asserts that the state of the system at time $t$ can be expanded in terms of $T$-periodic Floquet states $\ket{\phi_j(t)}$ and quasi-energies $\varepsilon_j\in [-\omega/2,\omega/2[$: 
\begin{equation}
	\ket{\psi(t)}	=\sum_{j=1}^N  \braket{\phi_j(t_0)|\psi(t_0)}   e^{-i\varepsilon_j (t-t_0)/\hbar}  \ket{\phi_j(t)}~.
\end{equation}
The Floquet states and quasi-energies are eigenvectors and eigenvalues of the Floquet Hamiltonian $H_\textrm{F} :=(H(t)-i\hbar\partial_t)$, considered as an operator on the Floquet-Hilbert space $\mathcal{H}_\textrm{F} := \mathcal{H} \otimes \mathcal{H}_T$ obtained by taking the tensor product of the original Hilbert space $\mathcal{H}$ with the space  of square-integrable $T$-periodic functions $\mathcal{H}_T$. 
Thus $H_\textrm{F} \ket{\phi_j(t)}=\varepsilon_j\ket{\phi_j(t)}$, where the time dependence is completely incorporated into the structure of the Floquet-Hilbert space~\cite{Sauer13}.

The Floquet formalism enables the formulation of design principles for enhanced transport in the periodically time-dependent network. 
First, consider a two-site system driven on resonance, i.e.\ where the driving frequency $\omega$ matches the energy difference between the two sites. In the rotating-wave approximation (RWA) \cite{CT08}, the system performs Rabi oscillations between the two levels, leading to perfect transfer.
Equivalently, in the Floquet picture, application of Shirley-Floquet perturbation theory \cite{Shirley63,Shirley65} shows that the degeneracy between the resonant states is lifted at first order in $\sigma_\mathrm{v}/\omega$, yielding a doublet of Floquet states
\begin{equation}\label{eq:ideal_floq}
\begin{split}
\ket{\delta_+(t)}&=\frac{1}{\sqrt{2}}\left( \ket{\textrm{in}}+ i  e^{i\omega t} \ket{\textrm{out}}\right) \, ,\\	 
\ket{\delta_-(t)}&=\frac{1}{\sqrt{2}}\left( i  e^{-i\omega t}\ket{\textrm{in}}+ \ket{\textrm{out}}\right) \, 
\end{split}
\end{equation}
(recall that $\ket{\textrm{in}} := \ket{1}$ and $\ket{\textrm{out}} := \ket{N}$).
These states lead to a sinusoidal transfer probability $P(t)$ with unit amplitude.
Note that although the states \eqref{eq:ideal_floq} themselves are of zeroth order in $\sigma_{\textrm{v}} /\omega$, the matrix elements of $H_\mathrm{v}$, of order $\sigma_\mathrm{v}$, are essential to obtain this spectral structure.

\begin{figure}
	\centering
	\includegraphics[width=0.95\linewidth]{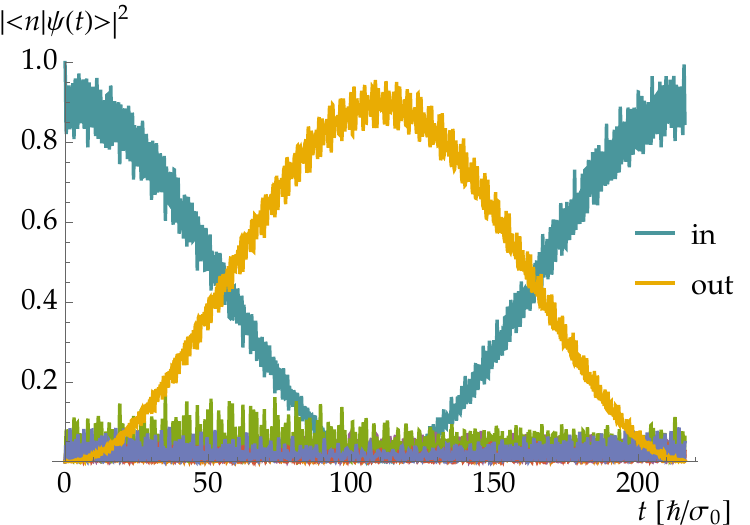}%
	\caption{\label{fig:1} Excitation transfer with a dominant Floquet doublet: Excitation probability on the input (blue), output (yellow) and  intermediate sites (other colors) for a randomly generated seven-site Floquet-antisymmetric network with the parameters $\sigma_\mathrm{d}=\sigma_0,\ D=10\, \sigma_0,\ \sigma_\mathrm{v}=0.1\, \sigma_0,\ \omega=\omega_0$ and satisfying the dominant Floquet doublet condition~\eqref{eq:dfd_cond} with $\beta\geq 0.9$.   }
\end{figure}

For an $N$-site network, we propose the \textit{dominant Floquet doublet} design principle for a time-periodic Hamiltonian at single-phonon resonance \cite{Kristjansson17}, which is satisfied if there exist two Floquet states $\ket{\phi_{\pm}(t)}$ which resemble the ideal states of Eq.\ \eqref{eq:ideal_floq}. To quantify the similarity of the two Floquet states, we use the inner product on the Floquet-Hilbert space, which involves an integral over a period of the driving to take into account the temporal dependence of the states. Thus, we say that the system has a  dominant Floquet doublet if
	\begin{equation}
		\label{eq:dfd_cond}
	\beta_\pm :=	\left|\frac{1}{T}\int_0^T\braket{\phi_{\pm}(t)|\delta_\pm(t)} \dif t \right|^2 \approx 1 \, . 
	\end{equation}

One can show that for a Hamiltonian satisfying the dominant Floquet doublet condition \eqref{eq:dfd_cond}, the transfer probability $P(t)$ essentially consists of a slow oscillation of large amplitude, 
\begin{equation}
\label{eq:appr_dfd_result}
P(t)\approx\frac{\beta_+^2+\beta_-^2}{4}  - \frac{\beta_+\beta_-}{2} \cos\left(\frac{\varepsilon_+-\varepsilon_-}{\hbar}(t-t_0)\right) \,,
\end{equation}
where $\varepsilon_\pm$ are the quasi-energies  associated with $\ket{\phi_\pm(t)}$. This oscillation is
garnished by small amplitude oscillations with frequencies of order $\omega$, as illustrated in Fig.~\ref{fig:1}. Note that intermediate sites also display small, rapidly oscillating, but non-zero populations. These are due to the finite contribution of Floquet states other than $\ket{\phi_\pm(t)}$ to the excitation amplitudes $\braket{n|\psi(t)}$.
A doublet strength $\beta :=\min(\beta_+,\beta_-)$ close to one therefore ensures efficient transfer, with a maximum transfer probability $P_\textrm{max} \ge \beta^2$ in the overwhelming number of cases, first reached at the \textit{transfer time} $\tau \approx \pi\hbar|\varepsilon_+-\varepsilon_-|^{-1}$.

\begin{figure}[htb]
	\centering
\begin{tabular}{l}
	(a)\\
	\includegraphics[width=0.9\linewidth]{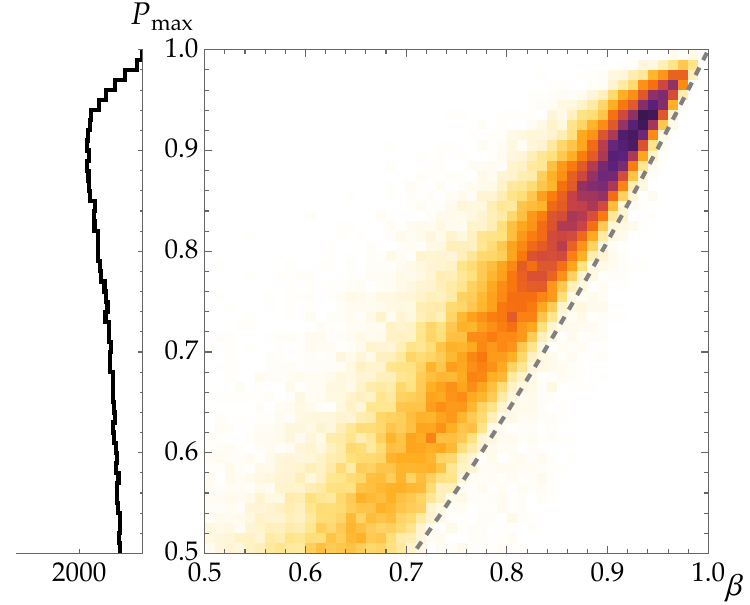}\\
	(b)\\
	\includegraphics[width=0.9\linewidth]{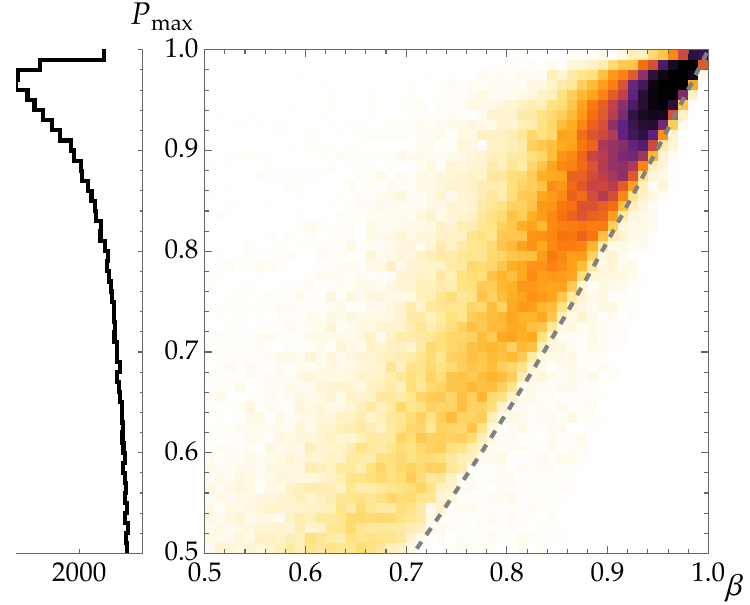}\\
	\hspace{0.25\linewidth} \includegraphics[width=0.5\linewidth]{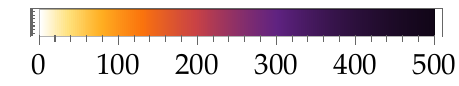} 
\end{tabular}
	\caption{\label{fig:2}Density histograms of $P_\textrm{max}$ vs.\ $\beta$ for $10^5$ randomly sampled seven-site Hamiltonians of the form~\eqref{eq:full_ham} with the parameters $\sigma_\mathrm{d}=\sigma_0,\ D=10\, \sigma_0,\ \sigma_\mathrm{v}=0.1\, \sigma_0,\ \omega=\omega_0$. Here, $P_\textrm{max}$ is taken as the maximum value of $P(t)$ within a time interval $2 \tau_{\rm exp}$, where $\tau_{\rm exp} = \pi\hbar|\varepsilon_+-\varepsilon_-|^{-1}$ is the expected  transfer time.
	In panel (a) no particular condition is imposed on the Hamiltonians  whereas in panel (b) the Hamiltonians are chosen to be Floquet-antisymmetric, i.e.\ $\{H_\textrm{F},\text{\emph{\TH}}\}=0$.
	The gray dashed line indicates the bound $P_\mathrm{\rm max} \geq \beta^2$. Only the realizations with $P_{\rm max}, \beta \geq 0.5$ are shown, of which there are
	55145 out of 100000 in (a) and 
	67654 out of 100000 in (b). The bin size is 0.01 in both axes. The marginal histograms of transfer probabilities are shown in the left panels.}
\end{figure}

We check the above analytical results numerically by randomly sampling $10^5$ seven-site Hamiltonians of the form of Eq.\ \eqref{eq:full_ham} and compute their Floquet eigenstates. For each Hamiltonian we evaluate the  doublet strength $\beta$, by considering the Floquet eigenstates with the highest overlap with $\ket{\delta_\pm (t)}$, as well as the maximum transfer probability $P_\textrm{max}$. The corresponding histogram is shown in Fig.~\ref{fig:2}(a), where the great majority of realizations are indeed found to lie above the line $P_\textrm{max}=\beta^2$.
The transport-enhancing doublet condition does not depend on any fine-tuning of the parameters, but is satisfied by some of the random realizations of the Hamiltonian.
This is analogous to dynamical tunneling between regular islands of a mixed regular-chaotic phase space \cite{davis1981quantum,lin1990quantum,peres1991instability,dyrting1993nonlinear}, which can be enhanced by the presence of the chaotic sea, leading to distinctive statistics of the transfer times and probabilities \cite{tomsovic1994chaos,Leyvarz96,Zakrzewski98,Erguen03}.

We now look for a symmetry-based design principle which enhances the realization of dominant Floquet doublets. The energy gradient of $H_\mathrm{d}$ in Eq.\ \eqref{eq:full_ham} suggests a reflection \emph{anti}symmetry, in contrast to the reflection symmetry found to enhance transport in static networks \cite{Walschaers13,Zech14}.
To consider reflection (anti)symmetry in the time degree of freedom, we introduce the time-reversal operator $\theta$ acting on functions $f\in \mathcal{H}_T$ as $\theta f(t) := f(T - t) = f(-t)$, where the last equality results from the $T$-periodicity of $f$. This operation can be seen as the analogue in time of the exchange operation $J$ in space. The formal similarity between the operators $J$ and $\theta$ motivates the definition of a generalized reversal operator on $\mathcal{H}_\textrm{F} = \mathcal{H}\otimes\mathcal{H}_T$ composed as a tensor product of $J$ and $\theta$. We denote this operator by the Icelandic letter `thorn', $\text{\emph{\TH}} := J \otimes \theta$, with its resulting action on a general state $\ket{\psi(t)} \in \mathcal{H}_\textrm{F}$ being $\text{\emph{\TH}} \ket{\psi(t)} = J \ket{\psi(-t)}$ \cite{Kristjansson17}.

We propose \textit{Floquet antisymmetry}, defined by the \emph{anti}commutation relation $\{H_\textrm{F},\text{\emph{\TH}}\}=0$, as a generalized symmetry-based design principle for the driven random network.
Given the anticommutation relations $\{-i\hbar \partial_t,\theta\}=0$ and $\{\sin(\omega t), \theta\}=0$, Floquet antisymmetry can be enforced by choosing $H_\textrm{v}$ centrosymmetric and $H_0$ and $H_\mathrm{d}$ anti-centrosymmetric, i.e.~$[H_\textrm{v},J]=\{H_0,J\}=\{H_\mathrm{d},J\}=0$ \cite{Kristjansson17}. 
In practice, random (anti-)centrosymmetric matrices are obtained by forming the combinations $(h\pm JhJ)/\sqrt{2}$, where $h$ is sampled from the corresponding unconstrained ensemble.

If $\ket{\phi(t)}$ is an eigenvector of a Floquet-antisymmetric operator $H_\textrm{F}$ with quasi-energy $\varepsilon$, then $\text{\emph{\TH}}\ket{\phi(t)}$ is also a Floquet state with quasi-energy $-\varepsilon$.  The ideal Floquet doublet states \eqref{eq:ideal_floq} form such a pair, with $\text{\emph{\TH}}\ket{\delta_+(t)}=\ket{\delta_-(t)}$. 
As a consequence, if a Floquet antisymmetric Hamiltonian has an eigenstate $\ket{\phi_+(t)}$ with large overlap with $\ket{\delta_+(t)}$, then there exists another state $\ket{\phi_-(t)}=\text{\emph{\TH}} \ket{\phi_+(t)}$ with equally large overlap with $\ket{\delta_-(t)}$. This means that the two conditions for a dominant Floquet doublet  [cf.\ Eq.\ \eqref{eq:dfd_cond}] reduce to only one. Hamiltonian realizations with a dominant Floquet doublet are therefore far more likely in a Floquet-antisymmetric ensemble than in a completely random one.

This is verified in numerical simulations, where the transport properties of $10^5$ Hamiltonians with Floquet antisymmetry are investigated. Fig.~\ref{fig:2}(b) shows the histogram of maximal transfer probabilities and doublet strengths. In comparison to the non-Floquet-antisymmetric case shown in Fig.~\ref{fig:2}(a), we observe a striking concentration of realizations with doublet strengths and transfer probabilities above 90\%
(note that the color scale is the same in both plots).

\emph{Two-phonon resonance.---} 
For networks with an odd number of sites $N$, Floquet-antisymmetry imposes that an eigenvalue of the static Hamiltonian $H_0+H_{\textrm{d}}$ lies exactly halfway between the highest and lowest eigenvalues.
Indeed, $H_0+H_{\textrm{d}}$ is anti-centrosymmetric, therefore its eigenvalues come in opposite pairs $e_n=-e_{N-n+1}$ and thus $e_{(N+1)/2}$ vanishes and  lies exactly between $e_1$ and $e_N=-e_1$.
This opens up the possibility of a resonant two-phonon transition from the input to the output site, in the presence of a single oscillation mode of frequency $\omega=\omega_0/2=e_1$.

\begin{figure}
	\centering
	\begin{tabular}{l}
		(a)\\
		\includegraphics[width=0.9\linewidth]{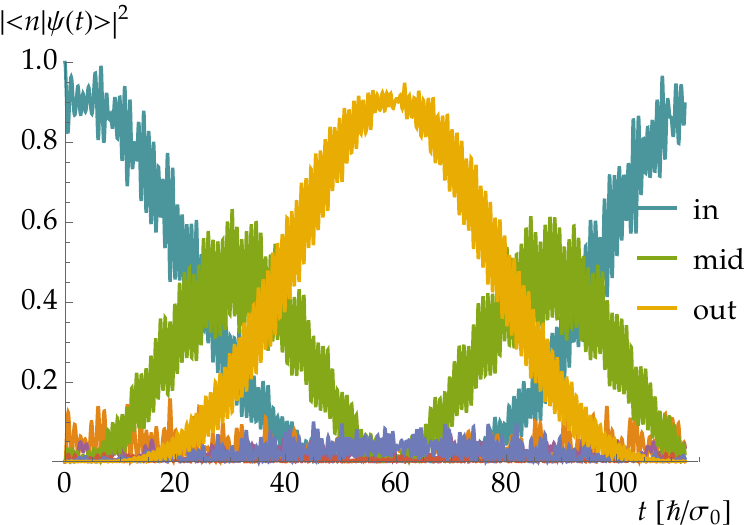}\\
		(b)\\
		\includegraphics[width=0.9\linewidth]{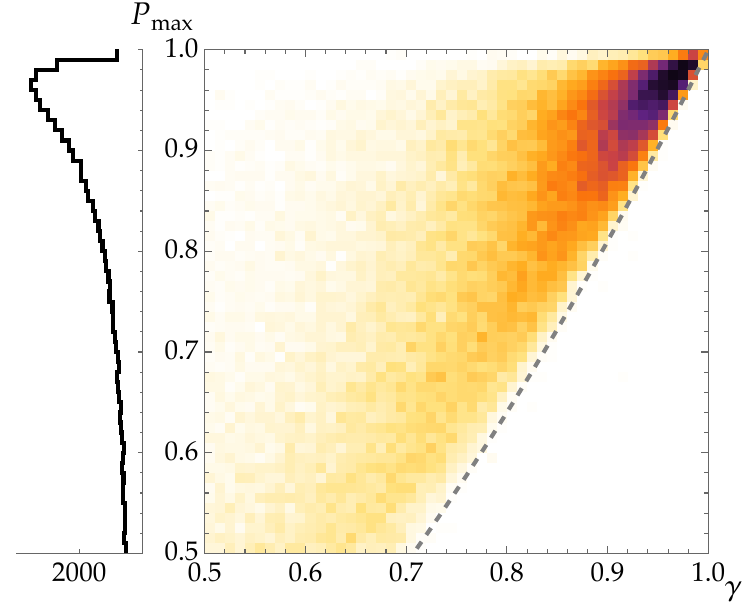}\\
		\hspace{0.25\linewidth} \includegraphics[width=0.5\linewidth]{fig_legend.pdf} 
	\end{tabular}
	\caption{\label{fig:3}Two-phonon transfer with a dominant Floquet triplet: (a) Excitation probability on the input (blue), output (yellow), central (green) and  other intermediate sites (other colors) for a randomly generated seven-site Floquet-antisymmetric network with the parameters $\sigma_\mathrm{d}=\sigma_0,\ D=10\, \sigma_0,\ \sigma_\mathrm{v}=0.1\, \sigma_0,\ \omega=\omega_0/2$, and satisfying the dominant Floquet triplet condition \eqref{eq:dft_cond} with $\gamma \geq 0.9$. (b) Density histogram of $P_\textrm{max}$ vs.\	$\gamma$ for $10^5$ randomly sampled Floquet antisymmetric Hamiltonians with the same parameters as in (a). Here, $P_\textrm{max}$ is taken as the maximum value of $P(t)$ within a time interval $2 \tau_{\rm exp}$, where $\tau_{\rm exp} = \pi\hbar| \varepsilon_+|^{-1}$ is the expected  transfer time. The gray dashed line indicates the bound $P_\mathrm{max} \geq \gamma^2$. Only realizations with $P_{\rm max}, \gamma \geq 0.5$ are shown, of which there are 64477
	out of the total 100000 sampled. The bin size is 0.01 in both axes. The marginal histogram of transfer probabilities is shown in the left panel.
 	}
\end{figure}

As in the case of single-phonon resonance, we can identify ideal Floquet states that ensure perfect transport, this time by considering a three-site network $\{\ket{\mathrm{in}}, \ket{\mathrm{mid}}, \ket{\mathrm{out}} \}$ with equally spaced on-site energies. The corresponding Floquet states in the RWA are
\begin{equation}
\label{eq:ideal_triplet}
\begin{split}
\ket{\tau_\pm(t)}&=\frac{1}{2}\left(\mp e^{-i\omega t } \ket{\mathrm{in}} + 
i\sqrt{2}\ket{\mathrm{mid}} \pm e^{i \omega t}\ket{\mathrm{out}} \right) \, , \\
\ket{\tau_0(t)}&=\frac{1}{\sqrt{2}}\left( e^{-i\omega t } \ket{\mathrm{in}} +e^{i \omega t}\ket{\mathrm{out}} \right) \, .
\end{split}
\end{equation}
Note that the two states $\ket{\tau_\pm(t)}$ are related by $\text{\emph{\TH}}$  while  $\ket{\tau_0(t)}$ is invariant under $\text{\emph{\TH}}$.

In an $N$-site network, the role of the $\ket{\mathrm{mid}}$ site can be played by any combination of intermediate sites. This means that in addition to $\ket{\tau_0}$, only the combination
\begin{equation}
	\begin{split}
\ket{\tau_1(t)} &:= \frac{1}{\sqrt{2}}(\ket{\tau_+(t)} - \ket{\tau_-(t)}) \\
&=\frac{1}{\sqrt{2}}\left( -e^{-i\omega t } \ket{\mathrm{in}} +e^{i \omega t}\ket{\mathrm{out}} \right) 
		\end{split}
\end{equation}
is relevant to describe transport from $\ket{\rm in}$ to $\ket{\rm out}$.
We then say that an $N$-site network has a \textit{dominant Floquet triplet} if the Hamiltonian has three Floquet eigenstates $\ket{\phi_{\pm}(t)}, \ket{\phi_{0}(t)}$ such that
\begin{equation}
\label{eq:dft_cond}
\gamma_{0,1} :=\left|\frac{1}{T}\int_0^T\braket{\phi_{0,1}(t)|\tau_{0,1}(t)} \dif t \right|^2\approx 1 \, , 
\end{equation}
where $\ket{\phi_1(t)} := (\ket{\phi_+(t)} - \ket{\phi_-(t)})/\sqrt{2}$.
If the Hamiltonian is Floquet-antisymmetric, i.e.\ $\{H_\textrm{F},\text{\emph{\TH}}\}=0$, then the quasi-energies associated with the triplet states obey $\varepsilon_++\varepsilon_-=\varepsilon_0=0$.

For a Floquet-antisymmetric Hamiltonian with a Floquet triplet strength $\gamma :=\min(\gamma_0,\gamma_1)\approx1$, close-to-perfect transport is achieved, with
\begin{equation}
\label{eq:appr_dft_result}
P(t) \approx \frac{1}{4}\left[ \gamma_0 - \gamma_1 \cos\left(\frac{\varepsilon_+ (t-t_0)}{\hbar} \right) \right]^2 \, ,
\end{equation}
Correspondingly, $P_\mathrm{max} \ge \gamma^2$ (in the vast majority of cases), with the maximum reached after a time $\pi\hbar |\varepsilon_+|^{-1}$.
A graph of the site occupation probabilities for a Floquet-antisymmetric Hamiltonian at two-phonon resonance is given in Fig.~\ref{fig:3}(a), where a significant population of one intermediate site can be observed. A numerical verification of the approximate bound $P_\mathrm{max} \ge \gamma^2$ is provided in Fig.~\ref{fig:3}(b).

\emph{Role of the static couplings.---} 
The preceding discussions may lead to the false impression that the transport is entirely controlled by the time-dependent part of the Hamiltonian and that the static couplings contained in $H_0$ play no role. 
We now show that, on the contrary, the static couplings are essential in determining both the transfer probability and the transfer time.
Indeed, vibrations do not directly drive transitions between the site basis states $\{\ket{n}\}$, but rather between the eigenstates $\{\ket{e_n}\}$ of the static Hamiltonian $H_0+H_{\textrm{d}}$ (which we will refer to as the `energy basis').
On the one hand, the static couplings cause a reduction in the transfer probability, because they rotate the eigenbasis $\{\ket{e_n}\}$ away from the site basis $\{\ket{n}\}$, meaning that exciting the lowest-energy state does not exactly correspond to the excitation being localised at the output site. 
On the other hand, the static couplings allow for a greater variability in the transfer times,
opening up the possibility of sizeable speed-ups with respect to the case where $H_0=0$.

\begin{figure}
	\centering
	\includegraphics[width=0.95\linewidth]{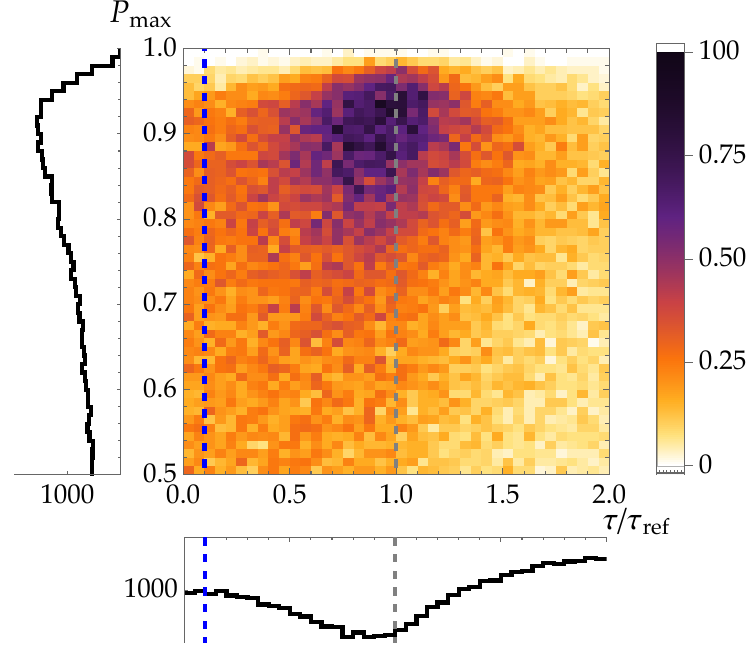}%
	\caption{\label{fig:transfertime} 
		Histogram of transfer probabilities $P_\mathrm{max}$ vs.\ transfer time enhancement $\tau/\tau_\mathrm{ref}$ for $10^5$ randomly generated seven-site networks with the parameters $\sigma_\mathrm{d}=\sigma_0,\ D=10\, \sigma_0,\ \sigma_\mathrm{v}=0.1\, \sigma_0,\ \omega=\omega_0$. Only the 46011 realizations for which  $\tau/\tau_{\rm ref} <2$ and  $P_{\rm max} > 0.5$ (of the the total 100000) are shown. The bin size is 0.01 for $P_{\rm max}$ and 0.05 for $\tau/\tau_{\rm ref}$. Realizations to the left of the dashed gray line, of which there are 28682, show an enhancement of the transfer time with respect to a network with no static couplings. Realizations to the left of the dashed blue line, of which there are 2066, show an order-of magnitude enhancement. The marginal histograms of transfer probabilities and transfer time enhancement are shown in the left and bottom panels, respectively.}
\end{figure}

The transfer times at one- and two-phonon resonance are approximately given by the inverse of the corresponding coupling matrix elements \emph{in the energy basis}, i.e.\ $\tau \approx \pi \hbar |\braket{e_1|H_\mathrm{v}|e_N}|^{-1}$  for single-phonon resonance, 
and  $ \tau \approx \sqrt{2} \pi\hbar |\braket{e_1|H_\mathrm{v}|e_{(N+1)/2}}|^{-1}$ 
in the (Floquet-antisymmetric) two-phonon case.
These values can differ significantly from those obtained from the bare couplings 
$\braket{\mathrm{in}|H_\mathrm{v}|\mathrm{out}}$ and $\braket{\mathrm{in}|H_\mathrm{v}|(N+1)/2}$. 
In Fig.~\ref{fig:transfertime}, we show a density histogram of transfer probability versus transfer time enhancement, i.e.\ the observed transfer time $\tau$ divided by the reference time $\tau_\mathrm{ref} \approx \pi \hbar |\braket{\mathrm{in}|H_\mathrm{v}|\mathrm{out}}|^{-1}$, for $10^5$ Hamiltonian realizations at one-phonon resonance.
The rotation from the site to the energy basis, at the cost of a moderate reduction of the transfer probability, can lead to an order-of-magnitude speed-up in transfer time, as evidenced by the region around $\tau/\tau_{\rm ref} = 0.1$ in Figure \ref{fig:transfertime}.

\emph{Conclusions.---} We have shown how near-perfect quantum transport in disordered networks with a large on-site energy gradient can be achieved by  coupling to a resonant vibrational mode. Design principles were introduced as constraints on the statistics, without any fine-tuning of individual parameters. Floquet antisymmetry, a composite reflection symmetry of the network in space and time, increases the probability of obtaining a dominant doublet of Floquet states. The network then behaves essentially like a resonantly driven two-level system and transport is almost perfect. However, the interplay between the static and vibrational  couplings allows some realizations to exhibit transport at timescales much faster than that which would occur in the absence of static couplings. This idea can be generalized to two-phonon transitions, and we envisage that, with the appropriate symmetries, efficient transitions involving a larger number of phonons may also be possible.

A more concrete application of our model to photosynthetic systems would require considering a broader spectrum of vibrations, the existence of loss and decoherence channels, as well as the mechanisms responsible for injecting and extracting the excitation from the network. Nevertheless, we anticipate that the principles described in this work can be applied to transport problems in a variety of fields, ranging from natural or artificial light-harvesting \cite{Romero17} to quantum state transfer \cite{Bose03,Christandl04, Keil13, Wojcik07}, for example in ultracold Rydberg gases \cite{Saffman10,Scholak14,Browaeys16}.

\emph{Acknowledgments.---}
We acknowledge fruitful discussions with Mattia Walschaers and Thomas Wellens. 
HK would like to thank the ERASMUS Programme for making this work possible, and the UK Engineering and Physical Sciences Research Council (EPSRC)  through grant EP/R513295/1, the Natural Sciences and Engineering Research Council of Canada (NSERC) through grant RGPIN-2025-04419, and support from the Institut Courtois, Faculté des arts et des sciences, Université de Montréal (Chaire Courtois de l’Institut Courtois) for subsequent research funding.
GD is grateful to the Alexander von Humboldt Foundation for its support. 



\begin{thebibliography}{53}%
	\makeatletter
	\providecommand \@ifxundefined [1]{%
		\@ifx{#1\undefined}
	}%
	\providecommand \@ifnum [1]{%
		\ifnum #1\expandafter \@firstoftwo
		\else \expandafter \@secondoftwo
		\fi
	}%
	\providecommand \@ifx [1]{%
		\ifx #1\expandafter \@firstoftwo
		\else \expandafter \@secondoftwo
		\fi
	}%
	\providecommand \natexlab [1]{#1}%
	\providecommand \enquote  [1]{``#1''}%
	\providecommand \bibnamefont  [1]{#1}%
	\providecommand \bibfnamefont [1]{#1}%
	\providecommand \citenamefont [1]{#1}%
	\providecommand \href@noop [0]{\@secondoftwo}%
	\providecommand \href [0]{\begingroup \@sanitize@url \@href}%
	\providecommand \@href[1]{\@@startlink{#1}\@@href}%
	\providecommand \@@href[1]{\endgroup#1\@@endlink}%
	\providecommand \@sanitize@url [0]{\catcode `\\12\catcode `\$12\catcode
		`\&12\catcode `\#12\catcode `\^12\catcode `\_12\catcode `\%12\relax}%
	\providecommand \@@startlink[1]{}%
	\providecommand \@@endlink[0]{}%
	\providecommand \url  [0]{\begingroup\@sanitize@url \@url }%
	\providecommand \@url [1]{\endgroup\@href {#1}{\urlprefix }}%
	\providecommand \urlprefix  [0]{URL }%
	\providecommand \Eprint [0]{\href }%
	\providecommand \doibase [0]{http://dx.doi.org/}%
	\providecommand \selectlanguage [0]{\@gobble}%
	\providecommand \bibinfo  [0]{\@secondoftwo}%
	\providecommand \bibfield  [0]{\@secondoftwo}%
	\providecommand \translation [1]{[#1]}%
	\providecommand \BibitemOpen [0]{}%
	\providecommand \bibitemStop [0]{}%
	\providecommand \bibitemNoStop [0]{.\EOS\space}%
	\providecommand \EOS [0]{\spacefactor3000\relax}%
	\providecommand \BibitemShut  [1]{\csname bibitem#1\endcsname}%
	\let\auto@bib@innerbib\@empty
	\bibitem [{\citenamefont {Bose}(2003)}]{Bose03}%
	\BibitemOpen
	\bibfield  {author} {\bibinfo {author} {\bibfnamefont {S.}~\bibnamefont
			{Bose}},\ }\href {\doibase 10.1103/PhysRevLett.91.207901} {\bibfield
		{journal} {\bibinfo  {journal} {Physical Review Letters}\ }\textbf {\bibinfo
			{volume} {91}},\ \bibinfo {pages} {207901} (\bibinfo {year}
		{2003})}\BibitemShut {NoStop}%
	\bibitem [{\citenamefont {Christandl}\ \emph {et~al.}(2004)\citenamefont
		{Christandl}, \citenamefont {Datta}, \citenamefont {Ekert},\ and\
		\citenamefont {Landahl}}]{Christandl04}%
	\BibitemOpen
	\bibfield  {author} {\bibinfo {author} {\bibfnamefont {M.}~\bibnamefont
			{Christandl}}, \bibinfo {author} {\bibfnamefont {N.}~\bibnamefont {Datta}},
		\bibinfo {author} {\bibfnamefont {A.}~\bibnamefont {Ekert}}, \ and\ \bibinfo
		{author} {\bibfnamefont {A.~J.}\ \bibnamefont {Landahl}},\ }\href {\doibase
		10.1103/PhysRevLett.92.187902} {\bibfield  {journal} {\bibinfo  {journal}
			{Phys. Rev. Lett.}\ }\textbf {\bibinfo {volume} {92}},\ \bibinfo {pages}
		{187902} (\bibinfo {year} {2004})}\BibitemShut {NoStop}%
	\bibitem [{\citenamefont {Perez-Leija}\ \emph {et~al.}(2013)\citenamefont
		{Perez-Leija}, \citenamefont {Keil}, \citenamefont {Kay}, \citenamefont
		{Moya-Cessa}, \citenamefont {Nolte}, \citenamefont {Kwek}, \citenamefont
		{Rodr\'{\i}guez-Lara}, \citenamefont {Szameit},\ and\ \citenamefont
		{Christodoulides}}]{Keil13}%
	\BibitemOpen
	\bibfield  {author} {\bibinfo {author} {\bibfnamefont {A.}~\bibnamefont
			{Perez-Leija}}, \bibinfo {author} {\bibfnamefont {R.}~\bibnamefont {Keil}},
		\bibinfo {author} {\bibfnamefont {A.}~\bibnamefont {Kay}}, \bibinfo {author}
		{\bibfnamefont {H.}~\bibnamefont {Moya-Cessa}}, \bibinfo {author}
		{\bibfnamefont {S.}~\bibnamefont {Nolte}}, \bibinfo {author} {\bibfnamefont
			{L.-C.}\ \bibnamefont {Kwek}}, \bibinfo {author} {\bibfnamefont {B.~M.}\
			\bibnamefont {Rodr\'{\i}guez-Lara}}, \bibinfo {author} {\bibfnamefont
			{A.}~\bibnamefont {Szameit}}, \ and\ \bibinfo {author} {\bibfnamefont
			{D.~N.}\ \bibnamefont {Christodoulides}},\ }\href {\doibase
		10.1103/PhysRevA.87.012309} {\bibfield  {journal} {\bibinfo  {journal} {Phys.
				Rev. A}\ }\textbf {\bibinfo {volume} {87}},\ \bibinfo {pages} {012309}
		(\bibinfo {year} {2013})}\BibitemShut {NoStop}%
	\bibitem [{\citenamefont {W\'ojcik}\ \emph {et~al.}(2007)\citenamefont
		{W\'ojcik}, \citenamefont {\L{}uczak}, \citenamefont
		{Kurzy\ifmmode~\acute{n}\else \'{n}\fi{}ski}, \citenamefont {Grudka},
		\citenamefont {Gdala},\ and\ \citenamefont {Bednarska}}]{Wojcik07}%
	\BibitemOpen
	\bibfield  {author} {\bibinfo {author} {\bibfnamefont {A.}~\bibnamefont
			{W\'ojcik}}, \bibinfo {author} {\bibfnamefont {T.}~\bibnamefont {\L{}uczak}},
		\bibinfo {author} {\bibfnamefont {P.}~\bibnamefont
			{Kurzy\ifmmode~\acute{n}\else \'{n}\fi{}ski}}, \bibinfo {author}
		{\bibfnamefont {A.}~\bibnamefont {Grudka}}, \bibinfo {author} {\bibfnamefont
			{T.}~\bibnamefont {Gdala}}, \ and\ \bibinfo {author} {\bibfnamefont
			{M.}~\bibnamefont {Bednarska}},\ }\href {\doibase 10.1103/PhysRevA.75.022330}
	{\bibfield  {journal} {\bibinfo  {journal} {Phys. Rev. A}\ }\textbf {\bibinfo
			{volume} {75}},\ \bibinfo {pages} {022330} (\bibinfo {year}
		{2007})}\BibitemShut {NoStop}%
	\bibitem [{\citenamefont {Anderson}(1958)}]{Anderson58}%
	\BibitemOpen
	\bibfield  {author} {\bibinfo {author} {\bibfnamefont {P.~W.}\ \bibnamefont
			{Anderson}},\ }\href {\doibase 10.1103/PhysRev.109.1492} {\bibfield
		{journal} {\bibinfo  {journal} {Phys. Rev.}\ }\textbf {\bibinfo {volume}
			{109}},\ \bibinfo {pages} {1492} (\bibinfo {year} {1958})}\BibitemShut
	{NoStop}%
	\bibitem [{\citenamefont {Chakraborty}\ \emph {et~al.}(2016)\citenamefont
		{Chakraborty}, \citenamefont {Novo}, \citenamefont {Ambainis},\ and\
		\citenamefont {Omar}}]{Chakraborty16}%
	\BibitemOpen
	\bibfield  {author} {\bibinfo {author} {\bibfnamefont {S.}~\bibnamefont
			{Chakraborty}}, \bibinfo {author} {\bibfnamefont {L.}~\bibnamefont {Novo}},
		\bibinfo {author} {\bibfnamefont {A.}~\bibnamefont {Ambainis}}, \ and\
		\bibinfo {author} {\bibfnamefont {Y.}~\bibnamefont {Omar}},\ }\href {\doibase
		10.1103/PhysRevLett.116.100501} {\bibfield  {journal} {\bibinfo  {journal}
			{Phys. Rev. Lett.}\ }\textbf {\bibinfo {volume} {116}},\ \bibinfo {pages}
		{100501} (\bibinfo {year} {2016})}\BibitemShut {NoStop}%
	\bibitem [{\citenamefont {Scholak}\ \emph {et~al.}(2010)\citenamefont
		{Scholak}, \citenamefont {Mintert}, \citenamefont {Wellens},\ and\
		\citenamefont {Buchleitner}}]{Scholak10}%
	\BibitemOpen
	\bibfield  {author} {\bibinfo {author} {\bibfnamefont {T.}~\bibnamefont
			{Scholak}}, \bibinfo {author} {\bibfnamefont {F.}~\bibnamefont {Mintert}},
		\bibinfo {author} {\bibfnamefont {T.}~\bibnamefont {Wellens}}, \ and\
		\bibinfo {author} {\bibfnamefont {A.}~\bibnamefont {Buchleitner}},\ }\enquote
	{\bibinfo {title} {Transport and entanglement},}\ in\ \href@noop {} {\emph
		{\bibinfo {booktitle} {Quantum Efficiency in Complex Systems, Part I:
				Biomolecular Systems, Vol. 83}}},\ \bibinfo {editor} {edited by\ \bibinfo
		{editor} {\bibfnamefont {E.~R.}\ \bibnamefont {Weber}}, \bibinfo {editor}
		{\bibfnamefont {M.}~\bibnamefont {Thorwart}}, \ and\ \bibinfo {editor}
		{\bibfnamefont {U.}~\bibnamefont {W\"urfel}}}\ (\bibinfo  {publisher}
	{Academic Press, San Diego},\ \bibinfo {year} {2010})\ pp.\ \bibinfo {pages}
	{1--38}\BibitemShut {NoStop}%
	\bibitem [{\citenamefont {Scholak}\ \emph {et~al.}(2011)\citenamefont
		{Scholak}, \citenamefont {de~Melo}, \citenamefont {Wellens}, \citenamefont
		{Mintert},\ and\ \citenamefont {Buchleitner}}]{Scholak11}%
	\BibitemOpen
	\bibfield  {author} {\bibinfo {author} {\bibfnamefont {T.}~\bibnamefont
			{Scholak}}, \bibinfo {author} {\bibfnamefont {F.}~\bibnamefont {de~Melo}},
		\bibinfo {author} {\bibfnamefont {T.}~\bibnamefont {Wellens}}, \bibinfo
		{author} {\bibfnamefont {F.}~\bibnamefont {Mintert}}, \ and\ \bibinfo
		{author} {\bibfnamefont {A.}~\bibnamefont {Buchleitner}},\ }\href {\doibase
		10.1103/PhysRevE.83.021912} {\bibfield  {journal} {\bibinfo  {journal} {Phys.
				Rev. E}\ }\textbf {\bibinfo {volume} {83}},\ \bibinfo {pages} {021912}
		(\bibinfo {year} {2011})}\BibitemShut {NoStop}%
	\bibitem [{\citenamefont {Walschaers}\ \emph {et~al.}(2013)\citenamefont
		{Walschaers}, \citenamefont {Diaz}, \citenamefont {Mulet},\ and\
		\citenamefont {Buchleitner}}]{Walschaers13}%
	\BibitemOpen
	\bibfield  {author} {\bibinfo {author} {\bibfnamefont {M.}~\bibnamefont
			{Walschaers}}, \bibinfo {author} {\bibfnamefont {J.~F.-d.-C.}\ \bibnamefont
			{Diaz}}, \bibinfo {author} {\bibfnamefont {R.}~\bibnamefont {Mulet}}, \ and\
		\bibinfo {author} {\bibfnamefont {A.}~\bibnamefont {Buchleitner}},\ }\href
	{\doibase 10.1103/PhysRevLett.111.180601} {\bibfield  {journal} {\bibinfo
			{journal} {Phys. Rev. Lett.}\ }\textbf {\bibinfo {volume} {111}},\ \bibinfo
		{pages} {180601} (\bibinfo {year} {2013})}\BibitemShut {NoStop}%
	\bibitem [{\citenamefont {Walschaers}\ \emph {et~al.}(2015)\citenamefont
		{Walschaers}, \citenamefont {Mulet}, \citenamefont {Wellens},\ and\
		\citenamefont {Buchleitner}}]{Walschaers15}%
	\BibitemOpen
	\bibfield  {author} {\bibinfo {author} {\bibfnamefont {M.}~\bibnamefont
			{Walschaers}}, \bibinfo {author} {\bibfnamefont {R.}~\bibnamefont {Mulet}},
		\bibinfo {author} {\bibfnamefont {T.}~\bibnamefont {Wellens}}, \ and\
		\bibinfo {author} {\bibfnamefont {A.}~\bibnamefont {Buchleitner}},\ }\href
	{\doibase 10.1103/PhysRevE.91.042137} {\bibfield  {journal} {\bibinfo
			{journal} {Phys. Rev. E}\ }\textbf {\bibinfo {volume} {91}},\ \bibinfo
		{pages} {042137} (\bibinfo {year} {2015})}\BibitemShut {NoStop}%
	\bibitem [{\citenamefont {Mohseni}\ \emph {et~al.}(2013)\citenamefont
		{Mohseni}, \citenamefont {Shabani}, \citenamefont {Lloyd}, \citenamefont
		{Omar},\ and\ \citenamefont {Rabitz}}]{Mohseni13}%
	\BibitemOpen
	\bibfield  {author} {\bibinfo {author} {\bibfnamefont {M.}~\bibnamefont
			{Mohseni}}, \bibinfo {author} {\bibfnamefont {A.}~\bibnamefont {Shabani}},
		\bibinfo {author} {\bibfnamefont {S.}~\bibnamefont {Lloyd}}, \bibinfo
		{author} {\bibfnamefont {Y.}~\bibnamefont {Omar}}, \ and\ \bibinfo {author}
		{\bibfnamefont {H.}~\bibnamefont {Rabitz}},\ }\href {\doibase
		10.1063/1.4807084} {\bibfield  {journal} {\bibinfo  {journal} {The Journal of
				Chemical Physics}\ }\textbf {\bibinfo {volume} {138}},\ \bibinfo {pages}
		{204309} (\bibinfo {year} {2013})}\BibitemShut {NoStop}%
	\bibitem [{\citenamefont {Schmidt~am Busch}\ \emph {et~al.}(2011)\citenamefont
		{Schmidt~am Busch}, \citenamefont {M\"uh}, \citenamefont {El-Amine~Madjet},\
		and\ \citenamefont {Renger}}]{SchmidtAmBusch11}%
	\BibitemOpen
	\bibfield  {author} {\bibinfo {author} {\bibfnamefont {M.}~\bibnamefont
			{Schmidt~am Busch}}, \bibinfo {author} {\bibfnamefont {F.}~\bibnamefont
			{M\"uh}}, \bibinfo {author} {\bibfnamefont {M.}~\bibnamefont
			{El-Amine~Madjet}}, \ and\ \bibinfo {author} {\bibfnamefont {T.}~\bibnamefont
			{Renger}},\ }\href {\doibase 10.1021/jz101541b} {\bibfield  {journal}
		{\bibinfo  {journal} {The Journal of Physical Chemistry Letters}\ }\textbf
		{\bibinfo {volume} {2}},\ \bibinfo {pages} {93} (\bibinfo {year}
		{2011})}\BibitemShut {NoStop}%
	\bibitem [{\citenamefont {Blankenship}(2002)}]{Blankenship02}%
	\BibitemOpen
	\bibfield  {author} {\bibinfo {author} {\bibfnamefont {R.~E.}\ \bibnamefont
			{Blankenship}},\ }\href@noop {} {\emph {\bibinfo {title} {Molecular
				mechanisms of photosynthesis}}}\ (\bibinfo  {publisher} {Blackwell Science,
		Oxford},\ \bibinfo {year} {2002})\BibitemShut {NoStop}%
	\bibitem [{\citenamefont {Engel}\ \emph {et~al.}(2007)\citenamefont {Engel},
		\citenamefont {Calhoun}, \citenamefont {Read}, \citenamefont {Ahn},
		\citenamefont {Mancal}, \citenamefont {Cheng}, \citenamefont {Blankenship},\
		and\ \citenamefont {Fleming}}]{Engel07}%
	\BibitemOpen
	\bibfield  {author} {\bibinfo {author} {\bibfnamefont {G.~S.}\ \bibnamefont
			{Engel}}, \bibinfo {author} {\bibfnamefont {T.~R.}\ \bibnamefont {Calhoun}},
		\bibinfo {author} {\bibfnamefont {E.~L.}\ \bibnamefont {Read}}, \bibinfo
		{author} {\bibfnamefont {T.-K.}\ \bibnamefont {Ahn}}, \bibinfo {author}
		{\bibfnamefont {T.}~\bibnamefont {Mancal}}, \bibinfo {author} {\bibfnamefont
			{Y.-C.}\ \bibnamefont {Cheng}}, \bibinfo {author} {\bibfnamefont {R.~E.}\
			\bibnamefont {Blankenship}}, \ and\ \bibinfo {author} {\bibfnamefont {G.~R.}\
			\bibnamefont {Fleming}},\ }\href@noop {} {\bibfield  {journal} {\bibinfo
			{journal} {Nature}\ }\textbf {\bibinfo {volume} {446}},\ \bibinfo {pages}
		{782} (\bibinfo {year} {2007})}\BibitemShut {NoStop}%
	\bibitem [{\citenamefont {Christensson}\ \emph {et~al.}(2012)\citenamefont
		{Christensson}, \citenamefont {Kauffmann}, \citenamefont {Pullerits},\ and\
		\citenamefont {Man\v{c}al}}]{Christensson12}%
	\BibitemOpen
	\bibfield  {author} {\bibinfo {author} {\bibfnamefont {N.}~\bibnamefont
			{Christensson}}, \bibinfo {author} {\bibfnamefont {H.~F.}\ \bibnamefont
			{Kauffmann}}, \bibinfo {author} {\bibfnamefont {T.}~\bibnamefont
			{Pullerits}}, \ and\ \bibinfo {author} {\bibfnamefont {T.}~\bibnamefont
			{Man\v{c}al}},\ }\href@noop {} {\bibfield  {journal} {\bibinfo  {journal}
			{The Journal of Physical Chemistry B}\ }\textbf {\bibinfo {volume} {116}},\
		\bibinfo {pages} {7449} (\bibinfo {year} {2012})}\BibitemShut {NoStop}%
	\bibitem [{\citenamefont {Ishizaki}\ and\ \citenamefont
		{Fleming}(2009)}]{Ishizaki09}%
	\BibitemOpen
	\bibfield  {author} {\bibinfo {author} {\bibfnamefont {A.}~\bibnamefont
			{Ishizaki}}\ and\ \bibinfo {author} {\bibfnamefont {G.~R.}\ \bibnamefont
			{Fleming}},\ }\href {\doibase 10.1073/pnas.0908989106} {\bibfield  {journal}
		{\bibinfo  {journal} {Proceedings of the National Academy of Sciences}\
		}\textbf {\bibinfo {volume} {106}},\ \bibinfo {pages} {17255} (\bibinfo
		{year} {2009})}\BibitemShut {NoStop}%
	\bibitem [{\citenamefont {Scholes}\ \emph {et~al.}(2017)\citenamefont
		{Scholes}, \citenamefont {Fleming}, \citenamefont {Chen}, \citenamefont
		{Aspuru-Guzik}, \citenamefont {Buchleitner}, \citenamefont {Coker},
		\citenamefont {Engel}, \citenamefont {van Grondelle}, \citenamefont
		{Ishizaki}, \citenamefont {Jonas}, \citenamefont {Lundeen}, \citenamefont
		{McCusker}, \citenamefont {Mukamel}, \citenamefont {Ogilvie}, \citenamefont
		{Olaya-Castro}, \citenamefont {Ratner}, \citenamefont {Spano}, \citenamefont
		{Whaley},\ and\ \citenamefont {Zhu}}]{Scholes17}%
	\BibitemOpen
	\bibfield  {author} {\bibinfo {author} {\bibfnamefont {G.~D.}\ \bibnamefont
			{Scholes}}, \bibinfo {author} {\bibfnamefont {G.~R.}\ \bibnamefont
			{Fleming}}, \bibinfo {author} {\bibfnamefont {L.~X.}\ \bibnamefont {Chen}},
		\bibinfo {author} {\bibfnamefont {A.}~\bibnamefont {Aspuru-Guzik}}, \bibinfo
		{author} {\bibfnamefont {A.}~\bibnamefont {Buchleitner}}, \bibinfo {author}
		{\bibfnamefont {D.~F.}\ \bibnamefont {Coker}}, \bibinfo {author}
		{\bibfnamefont {G.~S.}\ \bibnamefont {Engel}}, \bibinfo {author}
		{\bibfnamefont {R.}~\bibnamefont {van Grondelle}}, \bibinfo {author}
		{\bibfnamefont {A.}~\bibnamefont {Ishizaki}}, \bibinfo {author}
		{\bibfnamefont {D.~M.}\ \bibnamefont {Jonas}}, \bibinfo {author}
		{\bibfnamefont {J.~S.}\ \bibnamefont {Lundeen}}, \bibinfo {author}
		{\bibfnamefont {J.~K.}\ \bibnamefont {McCusker}}, \bibinfo {author}
		{\bibfnamefont {S.}~\bibnamefont {Mukamel}}, \bibinfo {author} {\bibfnamefont
			{J.~P.}\ \bibnamefont {Ogilvie}}, \bibinfo {author} {\bibfnamefont
			{A.}~\bibnamefont {Olaya-Castro}}, \bibinfo {author} {\bibfnamefont {M.~A.}\
			\bibnamefont {Ratner}}, \bibinfo {author} {\bibfnamefont {F.~C.}\
			\bibnamefont {Spano}}, \bibinfo {author} {\bibfnamefont {K.~B.}\ \bibnamefont
			{Whaley}}, \ and\ \bibinfo {author} {\bibfnamefont {X.}~\bibnamefont {Zhu}},\
	}\href@noop {} {\bibfield  {journal} {\bibinfo  {journal} {Nature}\ }\textbf
		{\bibinfo {volume} {543}},\ \bibinfo {pages} {647} (\bibinfo {year}
		{2017})}\BibitemShut {NoStop}%
	\bibitem [{\citenamefont {Dost\'al}\ \emph {et~al.}(2016)\citenamefont
		{Dost\'al}, \citenamefont {P\u{s}en\u{c}\'ik},\ and\ \citenamefont
		{Zigmantas}}]{Dostal16}%
	\BibitemOpen
	\bibfield  {author} {\bibinfo {author} {\bibfnamefont {J.}~\bibnamefont
			{Dost\'al}}, \bibinfo {author} {\bibfnamefont {J.}~\bibnamefont
			{P\u{s}en\u{c}\'ik}}, \ and\ \bibinfo {author} {\bibfnamefont
			{D.}~\bibnamefont {Zigmantas}},\ }\href@noop {} {\bibfield  {journal}
		{\bibinfo  {journal} {Nature Chemistry}\ }\textbf {\bibinfo {volume} {8}},\
		\bibinfo {pages} {705} (\bibinfo {year} {2016})}\BibitemShut {NoStop}%
	\bibitem [{\citenamefont {Duan}\ \emph {et~al.}(2017)\citenamefont {Duan},
		\citenamefont {Prokhorenko}, \citenamefont {Cogdell}, \citenamefont {Ashraf},
		\citenamefont {Stevens}, \citenamefont {Thorwart},\ and\ \citenamefont
		{Miller}}]{duan2017nature}%
	\BibitemOpen
	\bibfield  {author} {\bibinfo {author} {\bibfnamefont {H.-G.}\ \bibnamefont
			{Duan}}, \bibinfo {author} {\bibfnamefont {V.~I.}\ \bibnamefont
			{Prokhorenko}}, \bibinfo {author} {\bibfnamefont {R.~J.}\ \bibnamefont
			{Cogdell}}, \bibinfo {author} {\bibfnamefont {K.}~\bibnamefont {Ashraf}},
		\bibinfo {author} {\bibfnamefont {A.~L.}\ \bibnamefont {Stevens}}, \bibinfo
		{author} {\bibfnamefont {M.}~\bibnamefont {Thorwart}}, \ and\ \bibinfo
		{author} {\bibfnamefont {R.~D.}\ \bibnamefont {Miller}},\ }\href@noop {}
	{\bibfield  {journal} {\bibinfo  {journal} {Proceedings of the National
				Academy of Sciences}\ }\textbf {\bibinfo {volume} {114}},\ \bibinfo {pages}
		{8493} (\bibinfo {year} {2017})}\BibitemShut {NoStop}%
	\bibitem [{\citenamefont {Cao}\ \emph {et~al.}(2020)\citenamefont {Cao},
		\citenamefont {Cogdell}, \citenamefont {Coker}, \citenamefont {Duan},
		\citenamefont {Hauer}, \citenamefont {Kleinekath{\"o}fer}, \citenamefont
		{Jansen}, \citenamefont {Man{\v{c}}al}, \citenamefont {Miller}, \citenamefont
		{Ogilvie} \emph {et~al.}}]{cao2020quantum}%
	\BibitemOpen
	\bibfield  {author} {\bibinfo {author} {\bibfnamefont {J.}~\bibnamefont
			{Cao}}, \bibinfo {author} {\bibfnamefont {R.~J.}\ \bibnamefont {Cogdell}},
		\bibinfo {author} {\bibfnamefont {D.~F.}\ \bibnamefont {Coker}}, \bibinfo
		{author} {\bibfnamefont {H.-G.}\ \bibnamefont {Duan}}, \bibinfo {author}
		{\bibfnamefont {J.}~\bibnamefont {Hauer}}, \bibinfo {author} {\bibfnamefont
			{U.}~\bibnamefont {Kleinekath{\"o}fer}}, \bibinfo {author} {\bibfnamefont
			{T.~L.}\ \bibnamefont {Jansen}}, \bibinfo {author} {\bibfnamefont
			{T.}~\bibnamefont {Man{\v{c}}al}}, \bibinfo {author} {\bibfnamefont {R.~D.}\
			\bibnamefont {Miller}}, \bibinfo {author} {\bibfnamefont {J.~P.}\
			\bibnamefont {Ogilvie}},  \emph {et~al.},\ }\href@noop {} {\bibfield
		{journal} {\bibinfo  {journal} {Science Advances}\ }\textbf {\bibinfo
			{volume} {6}},\ \bibinfo {pages} {eaaz4888} (\bibinfo {year}
		{2020})}\BibitemShut {NoStop}%
	\bibitem [{\citenamefont {Zerah~Harush}\ and\ \citenamefont
		{Dubi}(2021)}]{zerah2021photosynthetic}%
	\BibitemOpen
	\bibfield  {author} {\bibinfo {author} {\bibfnamefont {E.}~\bibnamefont
			{Zerah~Harush}}\ and\ \bibinfo {author} {\bibfnamefont {Y.}~\bibnamefont
			{Dubi}},\ }\href@noop {} {\bibfield  {journal} {\bibinfo  {journal} {Science
				Advances}\ }\textbf {\bibinfo {volume} {7}},\ \bibinfo {pages} {eabc4631}
		(\bibinfo {year} {2021})}\BibitemShut {NoStop}%
	\bibitem [{\citenamefont {Higgins}\ \emph {et~al.}(2021)\citenamefont
		{Higgins}, \citenamefont {Lloyd}, \citenamefont {Sohail}, \citenamefont
		{Allodi}, \citenamefont {Otto}, \citenamefont {Saer}, \citenamefont {Wood},
		\citenamefont {Massey}, \citenamefont {Ting}, \citenamefont {Blankenship}
		\emph {et~al.}}]{higgins2021photosynthesis}%
	\BibitemOpen
	\bibfield  {author} {\bibinfo {author} {\bibfnamefont {J.~S.}\ \bibnamefont
			{Higgins}}, \bibinfo {author} {\bibfnamefont {L.~T.}\ \bibnamefont {Lloyd}},
		\bibinfo {author} {\bibfnamefont {S.~H.}\ \bibnamefont {Sohail}}, \bibinfo
		{author} {\bibfnamefont {M.~A.}\ \bibnamefont {Allodi}}, \bibinfo {author}
		{\bibfnamefont {J.~P.}\ \bibnamefont {Otto}}, \bibinfo {author}
		{\bibfnamefont {R.~G.}\ \bibnamefont {Saer}}, \bibinfo {author}
		{\bibfnamefont {R.~E.}\ \bibnamefont {Wood}}, \bibinfo {author}
		{\bibfnamefont {S.~C.}\ \bibnamefont {Massey}}, \bibinfo {author}
		{\bibfnamefont {P.-C.}\ \bibnamefont {Ting}}, \bibinfo {author}
		{\bibfnamefont {R.~E.}\ \bibnamefont {Blankenship}},  \emph {et~al.},\
	}\href@noop {} {\bibfield  {journal} {\bibinfo  {journal} {Proceedings of the
				National Academy of Sciences}\ }\textbf {\bibinfo {volume} {118}},\ \bibinfo
		{pages} {e2018240118} (\bibinfo {year} {2021})}\BibitemShut {NoStop}%
	\bibitem [{\citenamefont {Mattioni}\ \emph {et~al.}(2021)\citenamefont
		{Mattioni}, \citenamefont {Caycedo-Soler}, \citenamefont {Huelga},\ and\
		\citenamefont {Plenio}}]{mattioni2021design}%
	\BibitemOpen
	\bibfield  {author} {\bibinfo {author} {\bibfnamefont {A.}~\bibnamefont
			{Mattioni}}, \bibinfo {author} {\bibfnamefont {F.}~\bibnamefont
			{Caycedo-Soler}}, \bibinfo {author} {\bibfnamefont {S.~F.}\ \bibnamefont
			{Huelga}}, \ and\ \bibinfo {author} {\bibfnamefont {M.~B.}\ \bibnamefont
			{Plenio}},\ }\href@noop {} {\bibfield  {journal} {\bibinfo  {journal}
			{Physical Review X}\ }\textbf {\bibinfo {volume} {11}},\ \bibinfo {pages}
		{041003} (\bibinfo {year} {2021})}\BibitemShut {NoStop}%
	\bibitem [{\citenamefont {Lorenzoni}\ \emph {et~al.}(2025)\citenamefont
		{Lorenzoni}, \citenamefont {Lacroix}, \citenamefont {Lim}, \citenamefont
		{Tamascelli}, \citenamefont {Huelga},\ and\ \citenamefont
		{Plenio}}]{lorenzoni2025full}%
	\BibitemOpen
	\bibfield  {author} {\bibinfo {author} {\bibfnamefont {N.}~\bibnamefont
			{Lorenzoni}}, \bibinfo {author} {\bibfnamefont {T.}~\bibnamefont {Lacroix}},
		\bibinfo {author} {\bibfnamefont {J.}~\bibnamefont {Lim}}, \bibinfo {author}
		{\bibfnamefont {D.}~\bibnamefont {Tamascelli}}, \bibinfo {author}
		{\bibfnamefont {S.~F.}\ \bibnamefont {Huelga}}, \ and\ \bibinfo {author}
		{\bibfnamefont {M.~B.}\ \bibnamefont {Plenio}},\ }\href@noop {} {\bibfield
		{journal} {\bibinfo  {journal} {Science Advances}\ }\textbf {\bibinfo
			{volume} {11}},\ \bibinfo {pages} {eady6751} (\bibinfo {year}
		{2025})}\BibitemShut {NoStop}%
	\bibitem [{\citenamefont {Aghtar}\ \emph {et~al.}(2014)\citenamefont {Aghtar},
		\citenamefont {Str\"umpfer}, \citenamefont {Olbrich}, \citenamefont
		{Schulten},\ and\ \citenamefont {Kleinekath\"ofer}}]{Aghtar14}%
	\BibitemOpen
	\bibfield  {author} {\bibinfo {author} {\bibfnamefont {M.}~\bibnamefont
			{Aghtar}}, \bibinfo {author} {\bibfnamefont {J.}~\bibnamefont {Str\"umpfer}},
		\bibinfo {author} {\bibfnamefont {C.}~\bibnamefont {Olbrich}}, \bibinfo
		{author} {\bibfnamefont {K.}~\bibnamefont {Schulten}}, \ and\ \bibinfo
		{author} {\bibfnamefont {U.}~\bibnamefont {Kleinekath\"ofer}},\ }\href
	{\doibase 10.1021/jz501351p} {\bibfield  {journal} {\bibinfo  {journal} {The
				Journal of Physical Chemistry Letters}\ }\textbf {\bibinfo {volume} {5}},\
		\bibinfo {pages} {3131} (\bibinfo {year} {2014})}\BibitemShut {NoStop}%
	\bibitem [{\citenamefont {Irish}\ \emph {et~al.}(2014)\citenamefont {Irish},
		\citenamefont {G\'omez-Bombarelli},\ and\ \citenamefont {Lovett}}]{Irish14}%
	\BibitemOpen
	\bibfield  {author} {\bibinfo {author} {\bibfnamefont {E.~K.}\ \bibnamefont
			{Irish}}, \bibinfo {author} {\bibfnamefont {R.}~\bibnamefont
			{G\'omez-Bombarelli}}, \ and\ \bibinfo {author} {\bibfnamefont {B.~W.}\
			\bibnamefont {Lovett}},\ }\href {\doibase 10.1103/PhysRevA.90.012510}
	{\bibfield  {journal} {\bibinfo  {journal} {Phys. Rev. A}\ }\textbf {\bibinfo
			{volume} {90}},\ \bibinfo {pages} {012510} (\bibinfo {year}
		{2014})}\BibitemShut {NoStop}%
	\bibitem [{\citenamefont {Nalbach}\ \emph {et~al.}(2015)\citenamefont
		{Nalbach}, \citenamefont {Mujica-Martinez},\ and\ \citenamefont
		{Thorwart}}]{Nalbach15}%
	\BibitemOpen
	\bibfield  {author} {\bibinfo {author} {\bibfnamefont {P.}~\bibnamefont
			{Nalbach}}, \bibinfo {author} {\bibfnamefont {C.~A.}\ \bibnamefont
			{Mujica-Martinez}}, \ and\ \bibinfo {author} {\bibfnamefont {M.}~\bibnamefont
			{Thorwart}},\ }\href {\doibase 10.1103/PhysRevE.91.022706} {\bibfield
		{journal} {\bibinfo  {journal} {Phys. Rev. E}\ }\textbf {\bibinfo {volume}
			{91}},\ \bibinfo {pages} {022706} (\bibinfo {year} {2015})}\BibitemShut
	{NoStop}%
	\bibitem [{\citenamefont {Gorman}\ \emph {et~al.}(2018)\citenamefont {Gorman},
		\citenamefont {Hemmerling}, \citenamefont {Megidish}, \citenamefont
		{Moeller}, \citenamefont {Schindler}, \citenamefont {Sarovar},\ and\
		\citenamefont {Haeffner}}]{gorman2018engineering}%
	\BibitemOpen
	\bibfield  {author} {\bibinfo {author} {\bibfnamefont {D.~J.}\ \bibnamefont
			{Gorman}}, \bibinfo {author} {\bibfnamefont {B.}~\bibnamefont {Hemmerling}},
		\bibinfo {author} {\bibfnamefont {E.}~\bibnamefont {Megidish}}, \bibinfo
		{author} {\bibfnamefont {S.~A.}\ \bibnamefont {Moeller}}, \bibinfo {author}
		{\bibfnamefont {P.}~\bibnamefont {Schindler}}, \bibinfo {author}
		{\bibfnamefont {M.}~\bibnamefont {Sarovar}}, \ and\ \bibinfo {author}
		{\bibfnamefont {H.}~\bibnamefont {Haeffner}},\ }\href@noop {} {\bibfield
		{journal} {\bibinfo  {journal} {Physical Review X}\ }\textbf {\bibinfo
			{volume} {8}},\ \bibinfo {pages} {011038} (\bibinfo {year}
		{2018})}\BibitemShut {NoStop}%
	\bibitem [{\citenamefont {Li}\ \emph {et~al.}(2022)\citenamefont {Li},
		\citenamefont {Ko}, \citenamefont {Yang}, \citenamefont {Sarovar},\ and\
		\citenamefont {Whaley}}]{li2022interplay}%
	\BibitemOpen
	\bibfield  {author} {\bibinfo {author} {\bibfnamefont {Z.-Z.}\ \bibnamefont
			{Li}}, \bibinfo {author} {\bibfnamefont {L.}~\bibnamefont {Ko}}, \bibinfo
		{author} {\bibfnamefont {Z.}~\bibnamefont {Yang}}, \bibinfo {author}
		{\bibfnamefont {M.}~\bibnamefont {Sarovar}}, \ and\ \bibinfo {author}
		{\bibfnamefont {K.~B.}\ \bibnamefont {Whaley}},\ }\href@noop {} {\bibfield
		{journal} {\bibinfo  {journal} {New Journal of Physics}\ }\textbf {\bibinfo
			{volume} {24}},\ \bibinfo {pages} {033032} (\bibinfo {year}
		{2022})}\BibitemShut {NoStop}%
	\bibitem [{\citenamefont {Policht}\ \emph {et~al.}(2022)\citenamefont
		{Policht}, \citenamefont {Niedringhaus}, \citenamefont {Willow},
		\citenamefont {Laible}, \citenamefont {Bocian}, \citenamefont {Kirmaier},
		\citenamefont {Holten}, \citenamefont {Man{\v{c}}al},\ and\ \citenamefont
		{Ogilvie}}]{policht2022hidden}%
	\BibitemOpen
	\bibfield  {author} {\bibinfo {author} {\bibfnamefont {V.~R.}\ \bibnamefont
			{Policht}}, \bibinfo {author} {\bibfnamefont {A.}~\bibnamefont
			{Niedringhaus}}, \bibinfo {author} {\bibfnamefont {R.}~\bibnamefont
			{Willow}}, \bibinfo {author} {\bibfnamefont {P.~D.}\ \bibnamefont {Laible}},
		\bibinfo {author} {\bibfnamefont {D.~F.}\ \bibnamefont {Bocian}}, \bibinfo
		{author} {\bibfnamefont {C.}~\bibnamefont {Kirmaier}}, \bibinfo {author}
		{\bibfnamefont {D.}~\bibnamefont {Holten}}, \bibinfo {author} {\bibfnamefont
			{T.}~\bibnamefont {Man{\v{c}}al}}, \ and\ \bibinfo {author} {\bibfnamefont
			{J.~P.}\ \bibnamefont {Ogilvie}},\ }\href@noop {} {\bibfield  {journal}
		{\bibinfo  {journal} {Science advances}\ }\textbf {\bibinfo {volume} {8}},\
		\bibinfo {pages} {eabk0953} (\bibinfo {year} {2022})}\BibitemShut {NoStop}%
	\bibitem [{\citenamefont {Floquet}(1883)}]{Floquet}%
	\BibitemOpen
	\bibfield  {author} {\bibinfo {author} {\bibfnamefont {G.}~\bibnamefont
			{Floquet}},\ }\href {https://eudml.org/doc/80895} {\bibfield  {journal}
		{\bibinfo  {journal} {Annales scientifiques de l'\'Ecole Normale
				Sup\'erieure}\ }\textbf {\bibinfo {volume} {12}},\ \bibinfo {pages} {47}
		(\bibinfo {year} {1883})}\BibitemShut {NoStop}%
	\bibitem [{\citenamefont {Shirley}(1965)}]{Shirley65}%
	\BibitemOpen
	\bibfield  {author} {\bibinfo {author} {\bibfnamefont {J.~H.}\ \bibnamefont
			{Shirley}},\ }\href {\doibase 10.1103/PhysRev.138.B979} {\bibfield  {journal}
		{\bibinfo  {journal} {Physical Review}\ }\textbf {\bibinfo {volume} {138}},\
		\bibinfo {pages} {B979} (\bibinfo {year} {1965})}\BibitemShut {NoStop}%
	\bibitem [{\citenamefont {Zel'dovich}(1967)}]{Zeldovich67}%
	\BibitemOpen
	\bibfield  {author} {\bibinfo {author} {\bibfnamefont {Y.~A.}\ \bibnamefont
			{Zel'dovich}},\ }\href@noop {} {\bibfield  {journal} {\bibinfo  {journal}
			{Soviet Physics JETP}\ }\textbf {\bibinfo {volume} {24}},\ \bibinfo {pages}
		{1006} (\bibinfo {year} {1967})}\BibitemShut {NoStop}%
	\bibitem [{\citenamefont {Oka}\ and\ \citenamefont
		{Kitamura}(2019)}]{oka2019floquet}%
	\BibitemOpen
	\bibfield  {author} {\bibinfo {author} {\bibfnamefont {T.}~\bibnamefont
			{Oka}}\ and\ \bibinfo {author} {\bibfnamefont {S.}~\bibnamefont {Kitamura}},\
	}\href@noop {} {\bibfield  {journal} {\bibinfo  {journal} {Annual Review of
				Condensed Matter Physics}\ }\textbf {\bibinfo {volume} {10}},\ \bibinfo
		{pages} {387} (\bibinfo {year} {2019})}\BibitemShut {NoStop}%
	\bibitem [{\citenamefont {Mehta}(2004)}]{mehta2004random}%
	\BibitemOpen
	\bibfield  {author} {\bibinfo {author} {\bibfnamefont {M.~L.}\ \bibnamefont
			{Mehta}},\ }\href@noop {} {\emph {\bibinfo {title} {Random matrices}}}\
	(\bibinfo  {publisher} {Elsevier},\ \bibinfo {year} {2004})\BibitemShut
	{NoStop}%
	\bibitem [{\citenamefont {Cohen-Tannoudji}\ \emph {et~al.}(1998)\citenamefont
		{Cohen-Tannoudji}, \citenamefont {Dupont-Roc},\ and\ \citenamefont
		{Grynberg}}]{CT08}%
	\BibitemOpen
	\bibfield  {author} {\bibinfo {author} {\bibfnamefont {C.}~\bibnamefont
			{Cohen-Tannoudji}}, \bibinfo {author} {\bibfnamefont {J.}~\bibnamefont
			{Dupont-Roc}}, \ and\ \bibinfo {author} {\bibfnamefont {G.}~\bibnamefont
			{Grynberg}},\ }\href@noop {} {\emph {\bibinfo {title} {Atom--Photon
				Interactions: Basic Process and Applications}}}\ (\bibinfo  {publisher}
	{Wiley-VCH},\ \bibinfo {year} {1998})\BibitemShut {NoStop}%
	\bibitem [{\citenamefont {Walschaers}(2016)}]{Walschaers16}%
	\BibitemOpen
	\bibfield  {author} {\bibinfo {author} {\bibfnamefont {M.}~\bibnamefont
			{Walschaers}},\ }\href@noop {} {\emph {\bibinfo {title} {Efficient quantum
				transport \emph{(Dissertation)}}}}\ (\bibinfo  {publisher}
	{Albert-Ludwigs-Universit\"at Freiburg, KU Leuven},\ \bibinfo {year}
	{2016})\BibitemShut {NoStop}%
	\bibitem [{\citenamefont {Zech}\ \emph {et~al.}(2014)\citenamefont {Zech},
		\citenamefont {Mulet}, \citenamefont {Wellens},\ and\ \citenamefont
		{Buchleitner}}]{Zech14}%
	\BibitemOpen
	\bibfield  {author} {\bibinfo {author} {\bibfnamefont {T.}~\bibnamefont
			{Zech}}, \bibinfo {author} {\bibfnamefont {R.}~\bibnamefont {Mulet}},
		\bibinfo {author} {\bibfnamefont {T.}~\bibnamefont {Wellens}}, \ and\
		\bibinfo {author} {\bibfnamefont {A.}~\bibnamefont {Buchleitner}},\
	}\href@noop {} {\bibfield  {journal} {\bibinfo  {journal} {New Journal of
				Physics}\ }\textbf {\bibinfo {volume} {16}},\ \bibinfo {pages} {055002}
		(\bibinfo {year} {2014})}\BibitemShut {NoStop}%
	\bibitem [{\citenamefont {Sauer}(2013)}]{Sauer13}%
	\BibitemOpen
	\bibfield  {author} {\bibinfo {author} {\bibfnamefont {S.}~\bibnamefont
			{Sauer}},\ }\href@noop {} {\emph {\bibinfo {title} {Entanglement in
				periodically driven quantum systems \emph{(Dissertation)}}}}\ (\bibinfo
	{publisher} {Albert-Ludwigs-Universit\"at Freiburg},\ \bibinfo {year}
	{2013})\BibitemShut {NoStop}%
	\bibitem [{\citenamefont {Shirley}(1963)}]{Shirley63}%
	\BibitemOpen
	\bibfield  {author} {\bibinfo {author} {\bibfnamefont {J.~H.}\ \bibnamefont
			{Shirley}},\ }\href@noop {} {\emph {\bibinfo {title} {Interaction of a
				quantum system with a strong oscillating field \emph{(PhD thesis)}}}}\
	(\bibinfo  {publisher} {California Institute of Technology},\ \bibinfo {year}
	{1963})\BibitemShut {NoStop}%
	\bibitem [{\citenamefont {Kristj\'ansson}(2017)}]{Kristjansson17}%
	\BibitemOpen
	\bibfield  {author} {\bibinfo {author} {\bibfnamefont {H.}~\bibnamefont
			{Kristj\'ansson}},\ }\href@noop {} {\emph {\bibinfo {title} {Efficient
				quantum transport in disordered networks with vibrations \emph{({M}aster's
					thesis)}}}}\ (\bibinfo  {publisher} {Albert-Ludwigs-Universit\"at Freiburg \&
		Imperial College London},\ \bibinfo {year} {2017})\BibitemShut {NoStop}%
	\bibitem [{\citenamefont {Davis}\ and\ \citenamefont
		{Heller}(1981)}]{davis1981quantum}%
	\BibitemOpen
	\bibfield  {author} {\bibinfo {author} {\bibfnamefont {M.~J.}\ \bibnamefont
			{Davis}}\ and\ \bibinfo {author} {\bibfnamefont {E.~J.}\ \bibnamefont
			{Heller}},\ }\href@noop {} {\bibfield  {journal} {\bibinfo  {journal} {The
				Journal of Chemical Physics}\ }\textbf {\bibinfo {volume} {75}},\ \bibinfo
		{pages} {246} (\bibinfo {year} {1981})}\BibitemShut {NoStop}%
	\bibitem [{\citenamefont {Lin}\ and\ \citenamefont
		{Ballentine}(1990)}]{lin1990quantum}%
	\BibitemOpen
	\bibfield  {author} {\bibinfo {author} {\bibfnamefont {W.}~\bibnamefont
			{Lin}}\ and\ \bibinfo {author} {\bibfnamefont {L.}~\bibnamefont
			{Ballentine}},\ }\href@noop {} {\bibfield  {journal} {\bibinfo  {journal}
			{Phys. Rev. Lett.}\ }\textbf {\bibinfo {volume} {65}},\ \bibinfo {pages}
		{2927} (\bibinfo {year} {1990})}\BibitemShut {NoStop}%
	\bibitem [{\citenamefont {Peres}(1991)}]{peres1991instability}%
	\BibitemOpen
	\bibfield  {author} {\bibinfo {author} {\bibfnamefont {A.}~\bibnamefont
			{Peres}},\ }in\ \href@noop {} {\emph {\bibinfo {booktitle} {Quantum Chaos}}}\
	(\bibinfo  {publisher} {World Scientific},\ \bibinfo {year}
	{1991})\BibitemShut {NoStop}%
	\bibitem [{\citenamefont {Dyrting}\ \emph {et~al.}(1993)\citenamefont
		{Dyrting}, \citenamefont {Milburn},\ and\ \citenamefont
		{Holmes}}]{dyrting1993nonlinear}%
	\BibitemOpen
	\bibfield  {author} {\bibinfo {author} {\bibfnamefont {S.}~\bibnamefont
			{Dyrting}}, \bibinfo {author} {\bibfnamefont {G.}~\bibnamefont {Milburn}}, \
		and\ \bibinfo {author} {\bibfnamefont {C.}~\bibnamefont {Holmes}},\
	}\href@noop {} {\bibfield  {journal} {\bibinfo  {journal} {Phys. Rev. E}\
		}\textbf {\bibinfo {volume} {48}},\ \bibinfo {pages} {969} (\bibinfo {year}
		{1993})}\BibitemShut {NoStop}%
	\bibitem [{\citenamefont {Tomsovic}\ and\ \citenamefont
		{Ullmo}(1994)}]{tomsovic1994chaos}%
	\BibitemOpen
	\bibfield  {author} {\bibinfo {author} {\bibfnamefont {S.}~\bibnamefont
			{Tomsovic}}\ and\ \bibinfo {author} {\bibfnamefont {D.}~\bibnamefont
			{Ullmo}},\ }\href@noop {} {\bibfield  {journal} {\bibinfo  {journal} {Phys.
				Rev. E}\ }\textbf {\bibinfo {volume} {50}},\ \bibinfo {pages} {145} (\bibinfo
		{year} {1994})}\BibitemShut {NoStop}%
	\bibitem [{\citenamefont {Leyvraz}\ and\ \citenamefont
		{Ullmo}(1996)}]{Leyvarz96}%
	\BibitemOpen
	\bibfield  {author} {\bibinfo {author} {\bibfnamefont {F.}~\bibnamefont
			{Leyvraz}}\ and\ \bibinfo {author} {\bibfnamefont {D.}~\bibnamefont
			{Ullmo}},\ }\href {http://stacks.iop.org/0305-4470/29/i=10/a=030} {\bibfield
		{journal} {\bibinfo  {journal} {Journal of Physics A: Mathematical and
				General}\ }\textbf {\bibinfo {volume} {29}},\ \bibinfo {pages} {2529}
		(\bibinfo {year} {1996})}\BibitemShut {NoStop}%
	\bibitem [{\citenamefont {Zakrzewski}\ \emph {et~al.}(1998)\citenamefont
		{Zakrzewski}, \citenamefont {Delande},\ and\ \citenamefont
		{Buchleitner}}]{Zakrzewski98}%
	\BibitemOpen
	\bibfield  {author} {\bibinfo {author} {\bibfnamefont {J.}~\bibnamefont
			{Zakrzewski}}, \bibinfo {author} {\bibfnamefont {D.}~\bibnamefont {Delande}},
		\ and\ \bibinfo {author} {\bibfnamefont {A.}~\bibnamefont {Buchleitner}},\
	}\href@noop {} {\bibfield  {journal} {\bibinfo  {journal} {Phys. Rev. E}\
		}\textbf {\bibinfo {volume} {57}},\ \bibinfo {pages} {1458} (\bibinfo {year}
		{1998})}\BibitemShut {NoStop}%
	\bibitem [{\citenamefont {Erg\"un}\ and\ \citenamefont
		{Fyodorov}(2003)}]{Erguen03}%
	\BibitemOpen
	\bibfield  {author} {\bibinfo {author} {\bibfnamefont {G.}~\bibnamefont
			{Erg\"un}}\ and\ \bibinfo {author} {\bibfnamefont {Y.~V.}\ \bibnamefont
			{Fyodorov}},\ }\href {\doibase 10.1103/PhysRevE.68.046124} {\bibfield
		{journal} {\bibinfo  {journal} {Phys. Rev. E}\ }\textbf {\bibinfo {volume}
			{68}},\ \bibinfo {pages} {046124} (\bibinfo {year} {2003})}\BibitemShut
	{NoStop}%
	\bibitem [{\citenamefont {Romero}\ \emph {et~al.}(2017)\citenamefont {Romero},
		\citenamefont {Novoderezhkin},\ and\ \citenamefont {van
			Grondelle}}]{Romero17}%
	\BibitemOpen
	\bibfield  {author} {\bibinfo {author} {\bibfnamefont {E.}~\bibnamefont
			{Romero}}, \bibinfo {author} {\bibfnamefont {V.~I.}\ \bibnamefont
			{Novoderezhkin}}, \ and\ \bibinfo {author} {\bibfnamefont {R.}~\bibnamefont
			{van Grondelle}},\ }\href@noop {} {\bibfield  {journal} {\bibinfo  {journal}
			{Nature}\ }\textbf {\bibinfo {volume} {543}},\ \bibinfo {pages} {355}
		(\bibinfo {year} {2017})}\BibitemShut {NoStop}%
	\bibitem [{\citenamefont {Saffman}\ \emph {et~al.}(2010)\citenamefont
		{Saffman}, \citenamefont {Walker},\ and\ \citenamefont
		{M\o{}lmer}}]{Saffman10}%
	\BibitemOpen
	\bibfield  {author} {\bibinfo {author} {\bibfnamefont {M.}~\bibnamefont
			{Saffman}}, \bibinfo {author} {\bibfnamefont {T.~G.}\ \bibnamefont {Walker}},
		\ and\ \bibinfo {author} {\bibfnamefont {K.}~\bibnamefont {M\o{}lmer}},\
	}\href {\doibase 10.1103/RevModPhys.82.2313} {\bibfield  {journal} {\bibinfo
			{journal} {Rev. Mod. Phys.}\ }\textbf {\bibinfo {volume} {82}},\ \bibinfo
		{pages} {2313} (\bibinfo {year} {2010})}\BibitemShut {NoStop}%
	\bibitem [{\citenamefont {Scholak}\ \emph {et~al.}(2014)\citenamefont
		{Scholak}, \citenamefont {Wellens},\ and\ \citenamefont
		{Buchleitner}}]{Scholak14}%
	\BibitemOpen
	\bibfield  {author} {\bibinfo {author} {\bibfnamefont {T.}~\bibnamefont
			{Scholak}}, \bibinfo {author} {\bibfnamefont {T.}~\bibnamefont {Wellens}}, \
		and\ \bibinfo {author} {\bibfnamefont {A.}~\bibnamefont {Buchleitner}},\
	}\href {\doibase 10.1103/PhysRevA.90.063415} {\bibfield  {journal} {\bibinfo
			{journal} {Phys. Rev. A}\ }\textbf {\bibinfo {volume} {90}},\ \bibinfo
		{pages} {063415} (\bibinfo {year} {2014})}\BibitemShut {NoStop}%
	\bibitem [{\citenamefont {Browaeys}\ \emph {et~al.}(2016)\citenamefont
		{Browaeys}, \citenamefont {Barredo},\ and\ \citenamefont
		{Lahaye}}]{Browaeys16}%
	\BibitemOpen
	\bibfield  {author} {\bibinfo {author} {\bibfnamefont {A.}~\bibnamefont
			{Browaeys}}, \bibinfo {author} {\bibfnamefont {D.}~\bibnamefont {Barredo}}, \
		and\ \bibinfo {author} {\bibfnamefont {T.}~\bibnamefont {Lahaye}},\ }\href
	{http://stacks.iop.org/0953-4075/49/i=15/a=152001} {\bibfield  {journal}
		{\bibinfo  {journal} {Journal of Physics B: Atomic, Molecular and Optical
				Physics}\ }\textbf {\bibinfo {volume} {49}},\ \bibinfo {pages} {152001}
		(\bibinfo {year} {2016})}\BibitemShut {NoStop}%
\end{thebibliography}

%

\end{document}